\documentclass[
   aps,
   pra,
   twocolumn,
   english,
   superscriptaddress,
   floatfix
]{revtex4-2}

\usepackage{
    times,
    bbold,
    bm,
    dsfont,
    amsmath,
    amsthm,
    amssymb
}

\usepackage{float}

\usepackage{physics}

\usepackage{mhchem}

\usepackage{booktabs}

\usepackage{color}
\usepackage[usenames, dvipsnames]{xcolor}
\usepackage{xkcdcolors}

\usepackage{xspace}

\usepackage{hyperref}
\hypersetup{
    pdftitle={Spin-state energetics of heme-related models with the variational quantum eigensolver},
    pdfauthor={Unathi Skosana, Sthembiso Gumede, Mark Tame},
    citecolor=PineGreen,
    linkcolor=RedViolet,
    urlcolor=RedViolet,
    colorlinks=true,
    linktocpage=true,
    breaklinks=true,
    bookmarksopen=true,
    backref=true,
    linktoc=all,
}

\usepackage[newcommands]{ragged2e}
\usepackage{graphicx}
\graphicspath{{gfx/}}
\usepackage[export]{adjustbox}
\usepackage[justification=Justified,singlelinecheck=false,format=plain]{caption}
\usepackage[justification=Justified,singlelinecheck=false,format=plain]{subcaption}

\begin{document}

\title{Spin-state energetics of a heme-related model with the variational quantum eigensolver}
\author{Unathi Skosana}
\email{ukskosana@gmail.com}
\author{Sthembiso Gumede}
\author{Mark Tame}
\affiliation{Department of Physics, Stellenbosch University, Matieland 7602, South Africa}

\begin{abstract}
	\noindent
We present numerical calculations of the energetic separation between different spin states (singlet, triplet and quintet) for a simplified model of a deoxy-myoglobin protein using the variational quantum eigensolver (VQE) algorithm. The goal is to gain insight into the workflow and challenges of VQE simulations for transition metal complexes, with emphasis on methodology over hardware-specific implementation. The numerical calculations are performed using an in-house statevector simulator with single- and multi-reference trial wavefunctions based on the $k$-unitary pair coupled-cluster generalized singles and doubles or $k$-UpCCGSD ansatz. The spin-state energetics for active spaces of increasing size up to 10 spatial orbitals (20 spin orbitals or qubits) are computed with VQE and were found to agree with the classical complete active self-consistent field or CASSCF method to within 1-4 kcal/mol. We evaluate relevant multi-reference diagnostics and show that the spin states computed with VQE possess a sufficient degree of multi-reference character to highlight the presence of strong electron correlation effects. Our numerical simulations show that in the ideal case, the VQE algorithm is capable of reproducing spin-state energetics of strongly correlated systems such as transition metal complexes for both single- and multi-reference trial wavefunctions, asymptotically achieving good agreement with results from classical methods as the number of active orbitals increases.
\end{abstract}

\maketitle

\section{Introduction\label{sec:introduction}}

In transition metal chemistry, and consequently bioinorganic chemistry,
determining the energetic separation between different spin states of
transition metal complexes is a crucial step in understanding their magnetic
and spectroscopic properties, and their chemical
reactions~\cite{Radon2014,Radon2015}. The reactivity patterns of transition
metal complexes in biological processes, such as the transport and storage of
dioxygen with hemoglobin and myoglobin, can be rationalized from an analysis
that starts with accurately computing the relative energies between different
spin states, or spin-state energetics~\cite{Carreon-Macedo2004,Strickland2007}.
Moreover, spin-state energetics are used subsequently to determine the
energetically most favorable spin state and the ordering of different spin
states, which are known to change during the course of a reaction, e.g.,
spin-forbidden reactions with transition metal complexes such as the heme
group~\cite{Strickland2007}.

The spin-state energetics of the heme group have been studied with density
functional theory (DFT), which is by far the most commonly used computational
quantum chemistry method. The predictions made by DFT with different
exchange-correlation functionals for equilibrium geometries are calibrated
against X-ray crystal structure experimental data. However, the situation is
different for single-point energies, as there is little experimental data to
calibrate against~\cite{Strickland2007}. Given these limitations, the
reliability and accuracy of DFT calculations are assessed with correlated
\emph{ab initio} methods, such as coupled-cluster methods like coupled-cluster
singles and doubles with perturbative triples, or
CCSD(T)~\cite{Strickland2007,Radon2014,Radon2024}. Due to the size of
transition metal complexes such as the heme group, an assessment with
correlated \emph{ab initio} methods is often computationally prohibitive. For
this reason, the calibration is performed on small model systems which have
been shown to reproduce the spin-state energetics of their corresponding heme
compound~\cite{Radon2014}.

Single-reference coupled-cluster methods like CCSD(T) are considered to be
accurate and reliable for first-row mononuclear transition metal species, for
which the effects of static correlation are considered to be weak or moderate.
On the other hand, multi-reference approaches such as complete active space
(CAS) methods, e.g., complete active space self-consistent field
(CASSCF), complete active space configuration interaction (CASCI) and complete
active space perturbation theory (CASPT2), are well-established at
systematically accounting for static correlation effects in transition metal
complexes~\cite{Andersson1990,Andersson1992,Pierloot2003,Malmqvist2008,Vancoillie2011b}.
However, results obtained from CAS methods are dependent on
choosing an appropriate active space for the problem at hand, which can be
time-consuming and often based on personal experience. Techniques for an
automated construction of active spaces such as atomic valence active space
(AVAS)~\cite{Sayfutyarova2017}, automatic complete active space
(autoCAS)~\cite{Stein2019} and machine learning-based
approaches~\cite{Golub2021} make multi-reference calculations easier to
reproduce by non-experts and sidesteps the traditional approach.

\begin{figure*}[!ht]%
	\vspace{-3ex}
	\centering
	\begin{subfigure}[T]{0.27\linewidth}
		\caption{\label{fig:3qm5}}
		\raisebox{-4cm}{\includegraphics[width=0.9\linewidth]{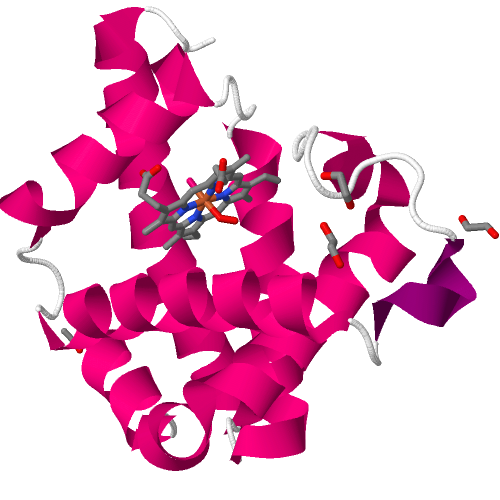}}
	\end{subfigure}
	\hfill
	\begin{subfigure}[T]{0.2\linewidth}
		\caption{\label{fig:fep_im_quintet}}
		\raisebox{-4cm}{\includegraphics[width=\linewidth]{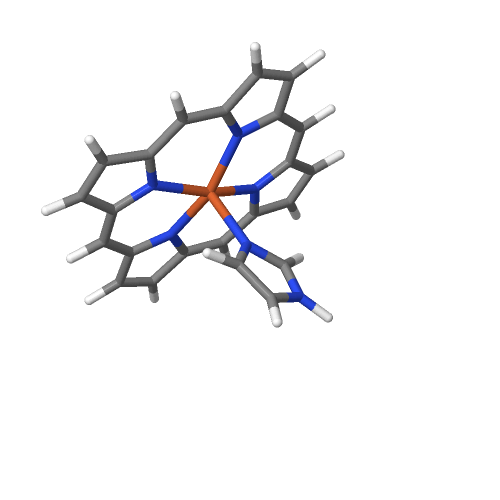}}
	\end{subfigure}
	\hfill
	\begin{subfigure}[T]{0.2\linewidth}
		\caption{\label{fig:model2_quintet}}
		\raisebox{-4cm}{\includegraphics[width=\linewidth]{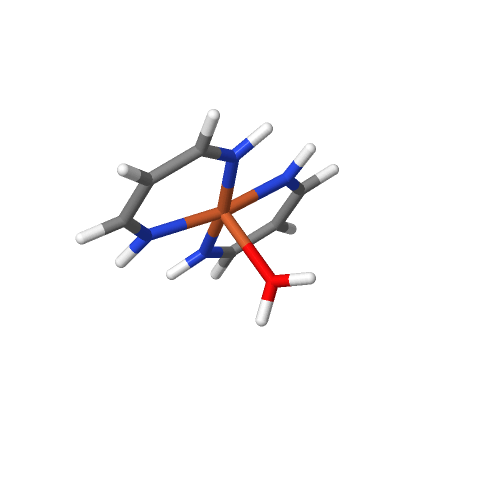}}
	\end{subfigure}
	\hfill
	\begin{subfigure}[T]{0.2\linewidth}
		\caption{\label{fig:model1_quintet}}
		\raisebox{-4cm}{\includegraphics[width=\linewidth]{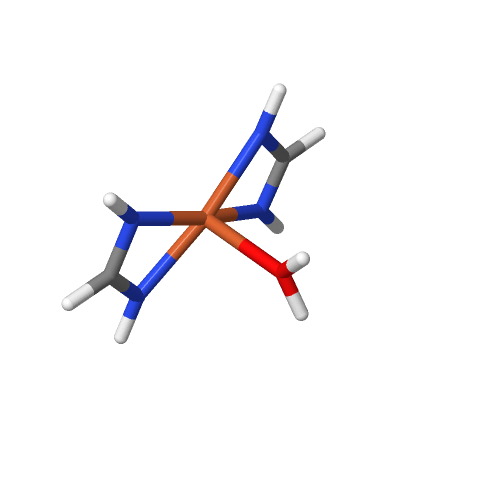}}
	\end{subfigure}
	\caption{
		Molecular structures of \textbf{(a)} Oxy-myoglobin in blackfin
		tuna (PDB \href{https://www.rcsb.org/structure/3qm5}{3QM5}),
		\textbf{(b)} \ce{FeP(Im)}, \textbf{(c)} \ce{Fe(C3H5N2)2(OH2)}
		and \textbf{(d)} \ce{Fe(CH3N2)2-(OH2)}, which is the transition
		metal complex studied in this work. Here, \textbf{(b)} is the
		most realistic model of the active site of deoxy-myoglobin
		(oxy-myoglobin without dioxygen) where the imidazole group from
		the side chain histidine is bound to the heme group, while
		\textbf{(c)} and \textbf{(d)} are model systems of
		\textbf{(b)}, with \textbf{(d)} being the most similar to
		\textbf{(b)}. Different colors correspond to different atoms.
		Here, red corresponds to oxygen (\ce{O}), blue to nitrogen
		(\ce{N}), gray to carbon (\ce{C}), and orange to iron
		(\ce{Fe}). The ribbons in \textbf{(a)} represent the
		surrounding polymers of the protein. The visualizations were
		made with PyMol and Jmol. See
		Appendix~\ref{sec:molecular_geometries} for more details about
		the structures.
		\label{fig:molecular_structures}
	}
\end{figure*}%

The feasibility of CAS methods is limited by the size of the active space, with
computational resources scaling exponentially with the number of active space
orbitals. However, classical tensor network algorithms, such as the Density
Matrix Renormalization Group (DMRG), can relax this scaling issue for systems
with moderate entanglement (a bond dimension value between 1000 and 10000 is
typically sufficient to achieve accurately converged energies), allowing one to
target much larger active spaces than traditional CAS methods~\cite{Chan2011}.
Quantum computing offers an alternative way of alleviating this scaling
behavior. The development of quantum algorithms for quantum chemistry opens up
the possibility to attain near-exact solutions of strongly correlated systems
using quantum computational resources that scale polynomially with the number
of active space orbitals~\cite{Cao2019,McArdle2020}. The application of quantum
computing to quantum chemistry problems is relatively nascent, with
state-of-the-art quantum hardware having to contend with hardware constraints
such as a limitation in the quantity and quality of available qubits, which
confines their scope of applicability.

Moreover, the development of quantum algorithms to work within the constraints
of near-term quantum hardware, such as the variational quantum eigensolver
(VQE)~\cite{Peruzzo2014,McClean2016}, and improvements to generally available
near-term quantum hardware have spurred active research in the realization of
complete active space calculations on quantum hardware. Despite the sound
theoretical guarantees of polynomial scaling and increasingly sophisticated
hardware experiments, potential issues still remain over VQE-based quantum
algorithms achieving any significant quantum advantage over classical
alternatives~\cite{Tilly2021}. These include a substantial number of
measurements required for accurate observable sampling, the exponential scaling
of gradient measurement resources due to the barren plateau (BP) problem, and resource
overhead from necessary error mitigation techniques. A detailed discussion of
these issues can be found in Ref.~\cite{Tilly2021}.

Nevertheless, progress in hardware and algorithmic development has led to a
growing body of empirical work demonstrating the feasibility of these
calculations on state-of-the-art quantum hardware. Table II in
Ref.~\cite{Nutzel2024} shows a list of CAS methods performed on quantum
hardware with VQE-based quantum algorithms. Only a handful of these studies
focus on transition metal complexes: the Fe(III)-NTA complex \ce{Fe(NTA)(H2O)2}
in Ref.~\cite{Nutzel2024}, and the iron-sulfur clusters
\ce{[Fe2S2(SCH3)4]^{2-}} and \ce{[Fe4S4(SCH3)4]^{2-}} in
Ref.~\cite{Robledomoreno2024}. According to the multi-reference diagnostics
reported in Ref.~\cite{Nutzel2024}; $T_{1}$ and $D_{1}$ multi-reference
diagnostics from coupled-cluster calculations~\cite{Lee1989,Janssen1998} and
the $Z_{s(1)}$ one-orbital entropy-based multi-reference
diagnostic~\cite{Stein2017}, the Fe(III)-NTA complex is claimed to be one of
the most complex quantum chemistry problems treated on quantum hardware to
date~\cite{Nutzel2024}. At the time of writing, this may no longer be the case
as more systems are being studied on quantum hardware that are not included in
the list in Ref.~\cite{Nutzel2024}. Notable exclusions include the triple-bond
breaking process in butyronitrile \ce{CH3CH2CH2CN}~\cite{Rossmannek2023}, and
hydration of \ce{CO2} in carbonic anhydrase enzymes~\cite{Ettenhuber2024}.

Besides quantum hardware experiments, as far as we are aware, only two other
works exist that quantitatively study via classical hardware
simulations, the potential use of a VQE-based quantum algorithm for modeling
transition metal complexes~\cite{Fitzpatrick2024,Farag2022}.
Ref.~\cite{Fitzpatrick2024} studies the spin-state energetics of ferrocene
\ce{Fe(C5H5)2} with a state-averaged ADAPT-VQE algorithm, termed ADAPT-VQE-SCF,
that uses a spin-preserving unitary coupled-cluster ansatz. An active space
with five orbitals was considered and an agreement up to $5$ kcal/mol was found
with their classical CASSCF reference data with seven orbitals, with the
difference attributed to the inclusion of two valence orbitals that were
previously in the core. In Ref.~\cite{Farag2022}, single-point ground state
energies of the transition metal complexes \ce{Li2CoO2} and \ce{Co2O4},
corresponding to the discharged and charged states of a lithium-ion battery,
were computed with a standard VQE algorithm using various unitary
coupled-cluster based ans\"{a}tze. Ref.~\cite{Farag2022} predicts ground state
energies that quantitatively agree with reference energies from coupled-cluster
singles and doubles (CCSD), but falls short at quantitatively reproducing the
reference energies from CASCI and CASSCF. This was attributed to the potential
lack of multi-reference character in the computed VQE wavefunctions.

Motivated by filling in the existing gap in the literature, and more
importantly the practical considerations of VQE-based methods in simulating
transition metal complexes, in this study we consider the accuracy of
simulations of the spin-state energetics for a simplified model of a
deoxy-myoglobin protein \ce{Fe(CH3N2)2-(OH2)} based on
Ref.~\cite{Strickland2007} (see Fig.~\ref{fig:molecular_structures}) with VQE.
To do this, we use a state-averaged orbital-optimization unitary
coupled-cluster based approach~\cite{Sokolov_2020,Mizukami_2020,Matousek2024},
in which all the spin states are computed simultaneously with a single set of
molecular orbitals and cluster operators. Here, state-averaging mitigates root
flipping, which occurs when states close in energy crossover during orbital
optimization~\cite{Seigbahn1981,Helgaker2014}, while the set of cluster
operators, if appropriately chosen, provide a way to target different spin
states without introducing undesired spin symmetry crossover during VQE
optimization~\cite{Greene-Diniz2020}. Using this approach, the spin-state
energetics are computed for active spaces identified with
AVAS~\cite{Sayfutyarova2017}, which automatically constructs molecular active
space orbitals from a set of molecular orbitals from a Hartree-Fock (HF) or
density functional theory (DFT) calculation and a target set of atomic valence
orbitals. Finally, we compute the one-orbital entropy-based multi-reference
diagnostic~\cite{Stein2017} and the orbital-pair mutual
information~\cite{Rissler2006,Boguslawski2013,Boguslawski2014} for the
different spin states in each active space as a way to access the
multi-reference character of the computed spin states. The goal of the work is
not to run VQE on quantum hardware, but to understand how to effectively
perform VQE simulations of transition metal complexes, and gain an
understanding of the workflow and challenges that arise.

The structure of the paper is organized as follows. In
Sec.~\ref{sec:preliminaries_and_methods}, we introduce and briefly summarize
the main tools used in this work and describe our methodology. Following this,
in Sec.~\ref{sec:results} we report our main results beginning with computed
spin-state energetics for the different active spaces and then followed by the
multi-reference diagnostics. Finally, in Sec.~\ref{sec:conclusion} we give
concluding remarks and discuss future work. An appendix is included.

\section{Preliminaries\label{sec:preliminaries_and_methods}}

\subsection{Molecular structures\label{sec:molecular_structures}}

Our calculations of spin energetics begin by determining the classically
computed benchmark energies and  equilibrium geometries of the simplified model
of a deoxy-myoglobin \ce{Fe(CH3N2)2-(OH2)}, as shown in
Fig.~\ref{fig:model1_quintet}, for each spin state. We follow the same
specifications as given in Ref.~\cite{Strickland2007}. The geometry
optimizations of the molecular structures for each spin state (singlet, triplet
and quintet) were carried out with Schr\"{o}dinger's Jaguar \emph{ab initio}
quantum chemistry software package (version 11.5, release 141), where we used
restricted open-shell Kohn-Sham (ROKS) DFT. The standard Los Alamos effective
core potential (ECP) with the Jaguar triple-$\zeta$ (LACV3P) basis set was used
for the metal center, and the 6-311G* basis set was used for the non-metal
atoms. The symmetry of the molecule was constrained to the $C_{2v}$ point
group, and all the atoms in the porphyrin ring (except for the iron) were
constrained to lie in a single plane to prevent distortions to the molecular
structure that decrease similarity with the full system (see
Fig.~\ref{fig:3qm5}). With the specifications above, we arrive at the result in
Tab.~\ref{tab:dft_abs_energies} and Tab.~\ref{tab:dft_rel_energies} in
Appendix~\ref{sec:molecular_geometries} for the absolute and relative energies
at the end of the geometry optimizations performed with different
exchange-correlation functionals, respectively (see Tab.~\ref{tab:jaguar_specs}
in Appendix~\ref{sec:molecular_geometries} for fine-grained specifications).
Both sets of results are in good agreement with Ref.~\cite{Strickland2007},
which were obtained with an older version of Jaguar software (version 6.0). See
Appendix~\ref{sec:molecular_geometries} for more details.

\subsection{Active spaces\label{sec:active_spaces}}

The automatic construction of active spaces for all spin states begins from the
B3LYP equilibrium geometry of \ce{Fe(CH3N2)2-(OH2)} for the quintet spin state
in Ref.~\cite{Strickland2007}. We chose the B3LYP equilibrium geometry as a
starting point primarily because we wish to compare our results against the
single-point CCSD(T) reference values for the spin-state energetics reported in
Ref.~\cite{Strickland2007}. The CCSD(T) method is highly regarded for its
accuracy in transition metal
chemistry~\cite{Harvey2011,Lawson2012,Hughes2012,Bross2013,Jiang2012}. From
this equilibrium geometry, we compute the molecular orbitals using a
symmetry-adapted restricted open-shell Hartree-Fock (ROHF) calculation in
PySCF~\cite{PySCF_2020}. Here, we made use of a composite
correlation-consistent basis set; cc-pVTZ basis set on the metal center and
cc-pVDZ basis set on the other atoms, in line with Ref.~\cite{Strickland2007}.
The symmetry-adapted ROHF molecular orbitals are used as input to AVAS
(available in PySCF) to construct active spaces of different sizes for
\ce{Fe(CH3N2)2-(OH2)}. The AVAS algorithm is classically efficient, as it scales
polynomially with the total number of molecular orbitals, i.e., $\sim
\mathcal{O}(N^3_\text{MO})$.

The size of the active space is varied by a numerical threshold parameter,
which measures the degree to which a molecular orbital overlaps with the space
spanned by a target set of atomic orbitals. Following the recommended
rule-of-thumb approach for selecting active spaces for transition metal
complexes based on active atomic orbitals~\cite{Veryazov2011}, we chose our
target set of atomic orbitals for AVAS as \ce{Fe} $3d$, \ce{Fe} $4d$ and \ce{O}
$2p_{z}$. We incrementally decrease the threshold parameter from 0.97 to 0.70
in such a way that the number of active orbitals increases by one, from 5 to 10
active orbitals. Since state averaging uses a single set of molecular orbitals
and cluster operators for all spin states, the active spaces for the singlet
and triplet spin states are appropriately constructed from the active space for
the quintet spin state in a straightforward manner (see
Appendix~\ref{sec:avas_active_spaces} for more details about the automatic
active space construction). This is in contrast to state-specific calculations,
where the active space orbitals are constructed independently using molecular
geometries optimized for each spin state. One of the issues that can arise with
state-specific calculations is that the active space orbitals for the different
spin states may not be qualitatively similar, e.g., orbitals with low occupancy
can rotate out of the active space into the virtual space for one spin state
but not for another~\cite{Radon2019}.

\subsection{Orbital optimized variational quantum eigensolver\label{sec:sa_vqe}}

The non-relativistic and spin-free molecular electronic Hamiltonian in the
absence of external fields in second quantization is given
by~\cite{Helgaker2014}
\begin{align}
	\hat{H}_{e} =  \sum_{ij} h_{ij} \hat{a}_{i}^{\dagger} \hat{a}_{j} +
	\frac{1}{2}\sum_{ijkl} g_{ijkl} \hat{a}_{i}^{\dagger} \hat{a}_{k}^{\dagger} \hat{a}_{l}  \hat{a}_{j} + \hat{V}_{nn},
	\label{eq:second_q_h_e}
\end{align}
where $\hat{a}_i^\dagger\>(\hat{a}_i)$ are fermionic creation (annihilation)
operators associated with a spin orbital $\chi_{i}$ (from a chosen basis set),
$\hat{V}_{nn}$ is the Coulomb repulsion between nuclei, and coefficients
$h_{ij}$ and $g_{ijkl}$ are molecular one- and two-electron integrals,
respectively. The indices $\{i,j,k,l\}$ label the one-particle spin orbitals.
In complete active space methods, one is interested in the active space
Hamiltonian given by
\begin{align}
	\hat{H}^{\text{CAS}}_{e} = \sum_{pq} \tilde{h}_{pq} \hat{a}^{\dagger}_{p}\hat{a}_{q} + \frac{1}{2} \sum_{pqrs} g_{pqrs} \hat{a}^{\dagger}_{p}\hat{a}^{\dagger}_{r}\hat{a}_{s}\hat{a}_{q},
	\label{eq:second_q_h_cas_e}
\end{align}
\noindent
where now the indices $p,q,r,s$ label only the active space orbitals, and the
one-electron integrals $h_{ij}$ are replaced by $\tilde{h}_{pq}$, which
accounts for the interactions between the active and inactive electrons, and
contains contributions from the one-electron integrals $h_{ij}$. The electronic
Hamiltonian in Eq.~\ref{eq:second_q_h_cas_e} is mapped to a qubit Hamiltonian
via the Jordan-Wigner mapping~\cite{Jordan_1928}, as implemented in
OpenFermion~\cite{McClean_2020}. See Ref.~\cite{Tilly2021} for a comprehensive
overview of Hamiltonian representation and fermion-to-qubit mappers. From this
mapping, $\hat{H}^{\text{CAS}}_{e}$ assumes the form
\begin{align}
	\hat{H}^{\text{CAS}}_{q} = \sum_{i} c_{i} \hat{\sigma}_{i}, \> \hat{\sigma}_{i} = \hat{o}^{(i)}_{1} \otimes \hat{o}^{(i)}_{2} \cdots \otimes \cdots,
\end{align}
\noindent
where the coefficients $c_i$ depend on $\tilde{h}_{pq}$ and $g_{pqrs}$ in
Eq.~\ref{eq:second_q_h_e}, and $\hat{\sigma}_{i}$ are tensor products of Pauli
operators $\hat{o}_{k}^{(i)} \in \{\mathbb{1}, \hat{\sigma}_{x},
\hat{\sigma}_{y}, \hat{\sigma}_{z}\}$. Since the mapping is isospectral,
$\hat{H}^{\text{CAS}}_{q}$ has the same spectrum as $\hat{H}^{\text{CAS}}_{e}$.
The variational quantum eigensolver (VQE) formulates finding an upper bound on
the total energy $E$ of the exact ground state of a molecule as a hybrid
quantum-classical variational optimization algorithm with respect to a set of
parameters $\boldsymbol{\theta}$ that parameterize a unitary operator $\hat{U}$
that acts on an appropriately initialized quantum state $\ket*{\Phi}$ of $N$
qubits, i.e.,
\begin{align}
	\ket*{\Phi(\boldsymbol{\theta})} = \hat{U}(\boldsymbol{\theta})\ket*{\Phi}.
	\label{eq:parameterization}
\end{align}
\noindent
For a fixed set of parameters, this is implemented as a quantum circuit and
the expectation value of $\hat{H}^{\text{CAS}}_{q}$ is evaluated from measurements of the
individual qubit operators $\hat{\sigma}_{i}$. The parameters $\boldsymbol{\theta}$
are iteratively learned by a classical algorithm to find a set that minimizes
the expectation value of $\hat{H}^{\text{CAS}}_{q}$, therefore
\begin{align}
	E & =
	\underset{\boldsymbol{\theta}}{\small\textsf{min}}\>\ev*{\hat{U}^{\dagger}(\boldsymbol{\theta})\hat{H}_{q}\hat{U}(\boldsymbol{\theta})}{\Phi},
	\nonumber \\ &= \underset{\boldsymbol{\theta}}{\small\textsf{min}} \sum_{i} c_{i}
	\ev*{\hat{U}^{\dagger}(\boldsymbol{\theta})\hat{\sigma}_{i}\hat{U}(\boldsymbol{\theta})}{\Phi}.
	\label{eq:vqe_exp_val}
\end{align}
\noindent
The variational optimized ground state energy $E$ provides an upper bound to
the exact ground state energy of $\hat{H}^{\text{CAS}}_{q}$. In addition to the
variational parameters $\boldsymbol{\theta}$, the molecular orbital basis
$\{\chi_{i}\}$ of Eq.~\ref{eq:second_q_h_cas_e} can also be variationally
optimized. It has been shown that UCC-based ans\"{a}tze lead to better results
when used in conjunction with orbital
optimization~\cite{Mizukami_2020,Sokolov_2020,Matousek2024}. During orbital
optimization a similarity transformation is applied to the second-quantized
Hamiltonian in Eq.~\ref{eq:second_q_h_cas_e} given by:
\begin{align}
	\hat{H}^{\text{CAS}}_{e} \to \hat{\tilde{H}}^{\text{CAS}}_{e} = e^{-\hat{\kappa}}\hat{H}^{\text{CAS}}_{e}e^{\hat{\kappa}},
\end{align}
\noindent
where $\hat{\kappa} = \sum_{pq}\kappa_{pq}(\hat{E}_{pq} - \hat{E}_{qp})$ is an
antihermitian operator. Here, $\hat{E}_{pq}$ is a single-excitation operator
\begin{align}
	\hat{E}_{pq} = \hat{a}^{\dagger}_{p,\alpha}\hat{a}_{q,\alpha} + \hat{a}^{\dagger}_{p,\beta}\hat{a}_{q,\beta},
\end{align}
\noindent
where $p,q$ denote the general molecular spatial orbital indices and
$\alpha,\beta$ denote the spin of the corresponding spatial orbital. This
similarity transformation with respect to $\hat{\kappa}$ is equivalent to a
rotation of the molecular orbital basis which $\hat{H}^{\text{CAS}}_{e}$ is
expanded in.  We will use the notation
\begin{align}
	E & =
	\underset{\boldsymbol{\theta},\boldsymbol{\kappa}}{\small\textsf{min}}\> \ev*{\hat{U}^{\dagger}(\boldsymbol{\theta})\hat{\tilde{H}}^{\text{CAS}}_{q}\hat{U}(\boldsymbol{\theta})}{\Phi},
	\label{eq:oo_vqe_exp_val}
\end{align}
\noindent
to denote a variational optimization of the total energy with respect to the
parameters $\boldsymbol{\theta}$ and orbital rotation parameters
$\boldsymbol{\kappa}$, where $\hat{\tilde{H}}^{\text{CAS}}_{q}$ is the qubit
Hamiltonian after applying the Jordan-Wigner mapping to
$\hat{\tilde{H}}^{\text{CAS}}_{e}$. The coupled optimization of the orbital and
variational parameters is solved via a second-order Newton-Raphson procedure,
where second derivatives couple $\boldsymbol{\kappa}$ to the first-order
correction of the two-particle density matrix $^1\Gamma^{ij}_{kl}$ approximated
via finite differences~\cite{Sun2017,Fitzpatrick2024}
\begin{align}
	^1 \Gamma^{ij}_{kl} \approx \frac{\Gamma^{ij}_{kl}[\boldsymbol{\theta}^{(s)} + \boldsymbol{\theta}^{(s+1)}] - \Gamma^{ij}_{kl}[\boldsymbol{\theta}^{(s)}]}{\boldsymbol{\theta}^{(s+1)} - \boldsymbol{\theta}^{(s)}},
\end{align}
\noindent
where $\boldsymbol{\theta}^{(s)}$ are variationally optimized parameters at
iteration $s$. The orbital rotation parameters $\kappa_{rs}$ are iteratively
learned via a classical optimizer as implemented in PySCF, which takes
as input the following one- and two-particle reduced density matrices computed
from the quantum state at the end of the variational optimization of the parameters
$\boldsymbol{\theta}$:
\begin{align}
	 & \gamma^{i}_{j} =
	\ev*{\hat{U}^{\dagger}(\boldsymbol{\theta})\hat{\sigma}^{+}_{i}\hat{\sigma}^{-}_{j}\hat{U}(\boldsymbol{\theta})}{\Phi},
	\nonumber           \\ &\Gamma^{ij}_{kl} =
	\ev*{\hat{U}^{\dagger}(\boldsymbol{\theta})\hat{\sigma}^{+}_{i}\hat{\sigma}^{+}_{j}\hat{\sigma}^{-}_{l}\hat{\sigma}^{-}_{k}\hat{U}(\boldsymbol{\theta})}{\Phi},
\end{align}
\noindent
where $\hat{\sigma}^{+}_{i} (\hat{\sigma}_{i}^{-})$ are qubit operators for
fermionic creation and annihilation operators $\hat{a}^{\dagger}_{i}
(\hat{a}_{i})$ after the Jordan-Wigner mapping. The VQE algorithm used in this
study is implemented using an in-house statevector simulator built on top of
JAX~\cite{Jax_2018} in order to take advantage of GPU computational resources.
Here, the variational parameter optimization of $\boldsymbol{\theta}$ uses the
Adaptive Moment Estimation (ADAM) optimization algorithm, as implemented in
Optax~\cite{Deepmind_2020}. We use default hyperparameters for ADAM but change
the default learning rate to use a polynomial schedule $f$
\begin{align}
	f(t) =
	\begin{cases}
		I,                                             & \text{if } t < B            \\
		(I - E)\left(1 - \frac{t - B}{T}\right)^P + E, & \text{if } B \leq t < B + T \\
		E,                                             & \text{if } t \geq B + T
	\end{cases}
\end{align}
\noindent
where $I=10^{-2},E=10^{-3},B=35000,T=10000$ and $P=2$. In this way, the
optimizer takes larger strides at the beginning and smaller strides as it
approaches the user-specified maximum optimization steps. The orbital
optimization is performed by PySCF's orbital optimizer, the entirety of the
CASSCF loop (VQE + orbital optimization) is orchestrated by PySCF and split
into different iterations (macro, micro and inner) which correspond to the
total number of CI, orbital and orbital rotation steps, respectively.

In the absence of external fields, a spin-free molecular Hamiltonian conserves
the electron number $\ev*{\hat{N}}$, the square of the total spin
$\ev*{\hat{S}^{2}}$, and $z$-component of the total spin $\ev*{\hat{S}_{z}}$
quantum numbers. However, the unconstrained energy optimization of the
corresponding qubit Hamiltonian in Eq.~\ref{eq:vqe_exp_val} does not
necessarily conserve all the aforementioned quantum
numbers~\cite{Ryabinkin2018}. The desired values for the quantum numbers
$\ev*{\hat{S}^{2}}, \ev*{\hat{S}_{z}}$ and $\ev*{\hat{N}}$ can be enforced in
two ways; adding penalty terms to Eq.~\ref{eq:vqe_exp_val} that penalize states
that do not have the desired quantum numbers~\cite{Ryabinkin2018}, or choosing a
quantum-number-preserving unitary $\hat{U}$ in Eq.~\ref{eq:parameterization}
that preserves some or all the quantum numbers~\cite{Anselmetti2021}. These two
approaches can be used together or separately. In this work, we use the latter
approach and choose a unitary $\hat{U}$ that preserves the total spin for each
of the spin states (singlet, triplet and quintet).

\subsection{Unitary coupled-cluster ans\"{a}tze\label{unitary_coupled_cluster_ansatze}}

For unitary coupled-cluster (UCC) based ans\"{a}tze~\cite{Anand2022}, the
parameterization in Eq.~\ref{eq:parameterization} takes the form
\begin{align}
	\ket*{\Phi(\boldsymbol{\theta})} = e^{\hat{T}(\boldsymbol{\theta}) - \hat{T}^{\dagger}(\boldsymbol{\theta})}\ket*{\Phi},
	\label{eq:ucc_ansatze_form}
\end{align}
\noindent
where the cluster operator $\hat{T}$ is a sum of qubit operators representing
fermionic single, double, etc., excitations after the Jordan-Wigner mapping.
Here, $\hat{T} - \hat{T}^{\dagger}$ is anti-hermitian and ensures the
exponentiation is a unitary operation. For the unitary coupled-cluster singles
and doubles (UCCSD) ansatz, the cluster operator $\hat{T}$ is truncated to a
sum of single and double excitations
\begin{align}
	 & \hat{T}(\boldsymbol{\theta}) = \hat{T}_{1}(\boldsymbol{\theta}) + \hat{T}_{2}(\boldsymbol{\theta}) \nonumber,                                               \\
	 & \hat{T}_{1}(\boldsymbol{\theta}) = \frac{1}{2}\sum_{pq} \theta^{q}_{p}\hat{\sigma}^{+}_{q}\hat{\sigma}^{-}_{p} \nonumber,                                   \\
	 & \hat{T}_{2}(\boldsymbol{\theta}) = \frac{1}{4}\sum_{pqrs}\theta_{pr}^{qs} \hat{\sigma}^{+}_{q}\hat{\sigma}^{+}_{s}\hat{\sigma}^{-}_{p}\hat{\sigma}^{-}_{r},
	\label{eq:uccsd_ansatze}
\end{align}
\noindent
where indices $q,s$ and  $p,r$ are restricted to unoccupied and occupied
orbitals, respectively. The cluster operator for the unitary coupled-cluster
generalized singles and doubles (UCCGSD) ansatz assumes the same form as
Eq.~\ref{eq:uccsd_ansatze}, however the indices $p,q,r,s$ are `generalized' and
make no distinction between occupied and unoccupied orbitals, allowing
occupied-occupied and unoccupied-unoccupied excitations in both $\hat{T}_{1}$
and $\hat{T}_{2}$. In the $k$-unitary pair coupled-cluster generalized singles
and doubles ($k$-UpCCGSD) ansatz, the cluster operator $\hat{T}$ includes
generalized single excitations and double excitations, which move electron
pairs between spatial orbitals. In contrast to Eq.~\ref{eq:ucc_ansatze_form},
the $k$-UpCCGSD cluster operator is applied $k$ times on the initial quantum
state:
\begin{align}
	\ket*{\Phi(\boldsymbol{\theta})} = \prod_{i=1}^{k}e^{\hat{T}^{(i)}(\boldsymbol{\theta}) - \hat{T}^{(i)^\dagger}(\boldsymbol{\theta})}\ket*{\Phi}.
\end{align}
\noindent
For each $i$, the parameters in $\hat{T}^{(i)}$ ($\theta_{p}^{q}$ and
$\theta_{pr}^{qs}$) are treated as independent during VQE optimization. In
comparison to other fixed structured UCC-based ans\"{a}tze, e.g., UCCSD and
UCCGSD, the $k$-UpCCGSD ansatz has a slower asymptotic growth rate in circuit
depth; scaling linearly with the number of spin orbitals (or number of
qubits)~\cite{Lee2018}. See Ref.~\cite{Anand2022} for an overview of unitary
coupled-cluster ans\"{a}tze. For all simulations presented here, the UCC-based
ans\"{a}tze are approximated by using a single Trotter step

\begin{align}
	e^{\hat{T}(\boldsymbol{\theta}) - \hat{T}^{\dagger}(\boldsymbol{\theta})} \approx \prod_{i} e^{\theta_i (\hat{g}_i - \hat{g}_i^{\dagger})},
	\label{eq:single_trotter_step}
\end{align}
\noindent
where $\hat{g}_i$ is a normal-ordered excitation operator. Despite this being
an approximation, the variational flexibility of the UCC ans\"{a}tze is
sufficient to offset the Trotter error~\cite{Grimsley2019}. Moreover, for the
purpose of speeding up the simulations, a single term in
Eq.~\ref{eq:single_trotter_step} is expanded into a polynomial form
as~\cite{Chen2020,Kottmann_2021,Rubin_2021}:
\begin{align}
	e^{\theta_i (\hat{g}_i - \hat{g}_i^{\dagger})} & = \mathbb{1} + \sin{\theta_{i}}(\hat{g}_i - \hat{g}_i^{\dagger})                                  \\
	                                               & \qquad +  (1 - \cos{\theta_i})(\hat{g}_i - \hat{g}_i^{\dagger})(\hat{g}_i - \hat{g}_i^{\dagger}).
	\nonumber
\end{align}
\noindent
For all simulations in this study, we use the $k$-UpCCGSD ansatz with $k=4$,
unless stated otherwise, because it uses fewer cluster operators (and hence
variational parameters) in comparison to other unitary coupled-cluster based
ans\"{a}tze~\cite{Lee2018}. Moreover, it has a variable number of Trotter steps
$k$ that can be adjusted to suit the available hardware. However, even with
these optimizations, UCC-based ans\"{a}tze including two-body (double
excitation) terms, such as $k$-UpCCGSD, are still susceptible to the barren
plateau (BP) phenomenon~\cite{McClean2018,Larocca2025}. Theoretical and
numerical evidence suggests that for such ans\"{a}tze, the variance of the cost
function gradient decays exponentially with the number of qubits ($n$) in the
system, even for small depths $k$~\cite{Mao2024}. This exponential vanish
implies that while $k$-UpCCGSD can be highly expressive, its trainability
quickly degrades as the active space size increases, imposing a strong
practical limit on the maximum number of qubits that can be effectively
simulated with such ans\"{a}tze~\cite{Mao2024}.

\subsection{State averaging\label{eq:state_averaging}}

The spin-state energetics in each active space are computed simultaneously with a
single set of molecular orbitals and UCC cluster operators by modifying the
energy functional in Eq.~\ref{eq:oo_vqe_exp_val} to a weighted average energy
functional:
\begin{align}
	E & =
	\underset{\boldsymbol{\theta},\boldsymbol{\kappa}}{\small \textsf{min}}\> \sum_{i} w_{i} \ev*{\hat{U}^{\dagger}(\boldsymbol{\theta})\hat{\tilde{H}}^{\text{CAS}}_{q}\hat{U}(\boldsymbol{\theta})}{\Phi_{i}},
	\label{eq:sa_oo_vqe_exp_val}
\end{align}
\noindent
where the fixed weights $\{w_i\}$ are chosen such that $\sum_{i} w_{i} = 1,
\forall w_i \in [0, 1]$; the most sensible choice being uniform weights. In our
case, we target the singlet, triplet and quintet spin states by initializing
the quantum states $\ket*{\Phi_{i}}$ with the appropriate quantum number for
$\ev*{\hat{S}^{2}}$, \emph{i.e.} $\ev*{\hat{S}^{2}} = 0,2,6$ for the singlet
($S = 0$), triplet ($S = 1$) and quintet ($S = 2$) spin states, respectively.
Optimizing over Eq.~\ref{eq:sa_oo_vqe_exp_val} using the same set of molecular
orbitals and cluster operators for all spin states ensures that all spin states
are treated on equal footing~\cite{Yalouz2021}. Additionally, all states that
start off as orthogonal remain orthogonal during a state averaged calculation,
mitigating crossover between the three spin states during the
optimization~\cite{PySCF_2020}. Practically, optimizing for three spin states
with the same set of molecular orbitals and cluster operators simultaneously is
computationally more efficient than optimizing for each spin state separately,
as in state-specific calculations.

\section{Results\label{sec:results}}

In this section, we report the main results of our study. We consider two types
of initial states for the different spin states $\{\ket*{\Phi_{i}}\}$ of
\ce{Fe(CH3N2)2-(OH2)} for each active space considered. The first type consists
of single-reference initial states in the VQE algorithm for all the spin
states. The second type consists of single-reference initial states for the
singlet and quintet spin states, while the initial state for the triplet spin
state is multi-reference and made up of a linear combination of two
single-reference states. We refer to the first and second type of initial state
configurations for the VQE algorithm as the T0 and T1 initial states,
respectively. Both types are appropriately constructed to have the desired
values for $\ev*{\hat{S}^{2}}$ and $\ev*{\hat{N}}$ (see
Appendix~\ref{sec:initial_states} for more details about the construction of
the initial states). The initial state for the triplet spin state being the
only multi-reference state was informed by preliminary state-specific VQE tests
where the triplet spin-state energy showed the largest deviations from CASSCF
when a single-reference state was used in comparison to the singlet and quintet
states under the same conditions. This observation suggested that the triplet
state's electronic structure was potentially the most challenging for the
single-reference starting point within this model system and might therefore
benefit from an initial state incorporating static correlation effects.

All computations in this work (VQE simulations and classical benchmarks) were
executed on an Intel Xeon Gold 5218 64-core CPU with a NVIDIA RTX A6000 PCIe 48
GB GPU and 502 GB of DDR4 RAM or an Intel(R) Xeon(R) Gold 6426Y 64-core CPU
with a NVIDIA Ada L40 PCIe 48 GB GPU and 252 GB DDR4 RAM. The numerical
precision was set to 64-bit floating point precision, unless stated otherwise.

\begin{figure*}[htp]
	\centering
	\begin{subfigure}[t]{0.49\linewidth}
		\setcounter{subfigure}{0}
		\raggedright
		\caption{\label{fig:t0_spin_energetics}}
		\includegraphics[width=0.90\linewidth]{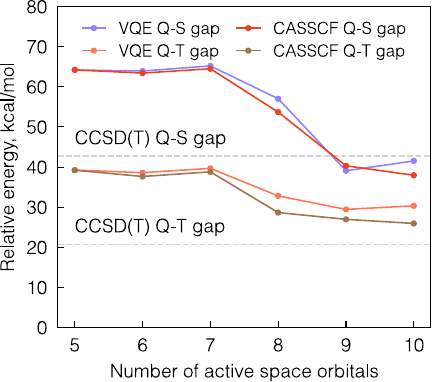}
	\end{subfigure}
	\begin{subfigure}[t]{0.49\linewidth}
		\setcounter{subfigure}{3}
		\raggedleft
		\caption{\label{fig:t1_spin_energetics}}
		\includegraphics[width=0.90\linewidth]{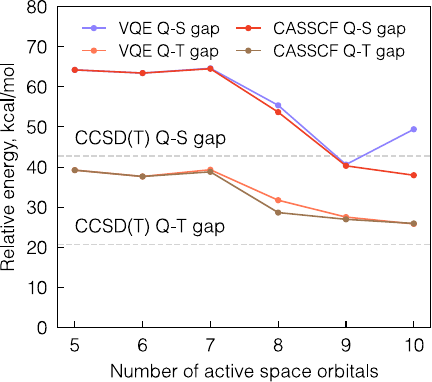}
	\end{subfigure}
	\begin{subfigure}[t]{0.49\linewidth}
		\setcounter{subfigure}{1}
		\raggedright
		\caption{\label{fig:t0_spin_energetics_error}}
		\includegraphics[width=0.90\linewidth]{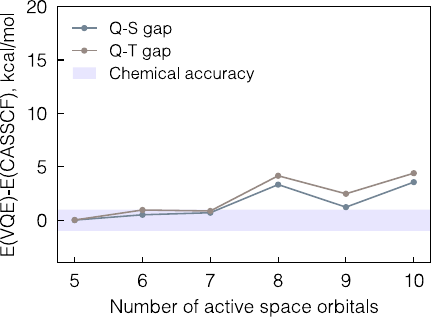}
	\end{subfigure}
	\begin{subfigure}[t]{0.49\linewidth}
		\setcounter{subfigure}{4}
		\raggedleft
		\caption{\label{fig:t1_spin_energetics_error}}
		\includegraphics[width=0.90\linewidth]{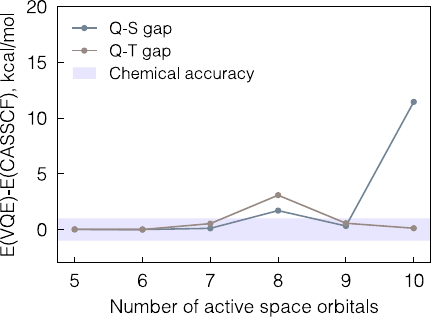}
	\end{subfigure}
	\begin{subfigure}[t]{0.49\linewidth}
		\setcounter{subfigure}{2}
		\raggedright
		\caption{\label{fig:t0_relative_energies}}
		\includegraphics[width=0.90\linewidth]{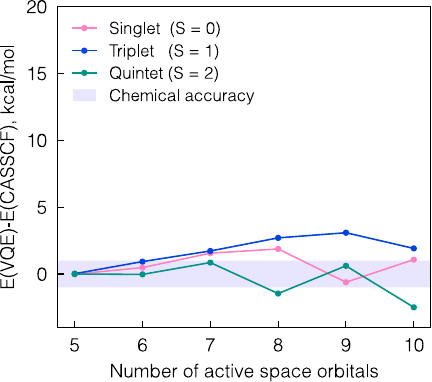}
	\end{subfigure}
	\begin{subfigure}[t]{0.49\linewidth}
		\setcounter{subfigure}{5}
		\raggedleft
		\caption{\label{fig:t1_relative_energies}}
		\includegraphics[width=0.90\linewidth]{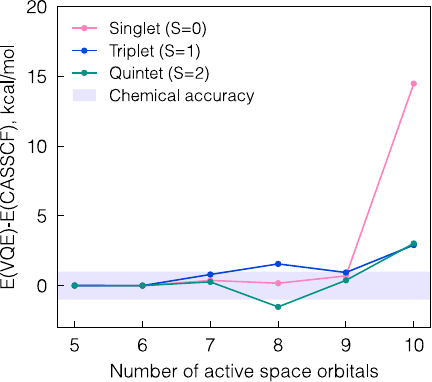}
	\end{subfigure}
	\caption{
		Spin-state energetics of \ce{Fe(CH3N2)2-(OH2)} as a function of
		the number of active space orbitals for the \textbf{(a)} T0 and
		\textbf{(d)} T1 initial states (dashed lines show reference
		CCSD(T) values from Ref.~\cite{Strickland2007}). Energy
		differences between the VQE and CASSCF spin-state energetics of
		\ce{Fe(CH3N2)2-(OH2)} as a function of the number of active
		space orbitals for the \textbf{(b)} T0 and \textbf{(e)} T1
		initial states. Energy differences between VQE and CASSCF for
		the individual spin states of \ce{Fe(CH3N2)2-(OH2)} as a
		function of the number of active space orbitals for the
		\textbf{(c)} T0 and \textbf{(f)} T1 initial states. See
		Appendix~\ref{sec:energy_traces} for energy traces during the
		optimization and Appendix~\ref{sec:active_space_total_energies}
		for the absolute/total energies for each active space.
		\label{fig:spin_energetics}
	}
\end{figure*}

\subsection{Spin energetics\label{sec:spin_energetics}}

We start by showing our results for the spin energetics computed with VQE and
compare them against those computed with the classical CASSCF method for the
same active space. Fig.~\ref{fig:t0_spin_energetics} shows the spin energetics
for the T0 initial states represented by the relative energy difference of the
singlet and triplet with respect to the quintet ground state.
Fig.~\ref{fig:t0_spin_energetics_error} shows the error energy differences
between the relative energies for VQE and CASSCF, and
Fig.~\ref{fig:t0_relative_energies} shows energy differences between the
individual spin states for VQE and CASSCF. Here, we observe for the active
spaces with 5 and 6 orbitals, the individual spin-state energies are in
agreement to within chemical accuracy ($\pm 1$ kcal/mol). As the active spaces
grow in size from 7 to 9 orbitals, there is still good agreement, but we start
to see a deviation reaching up to $\sim 3$ kcal/mol.

The spin energetics (energies relative to the quintet spin-state energy) follow
a similar trend (Fig.~\ref{fig:t0_spin_energetics}). For the small active
spaces with 5 to 7 orbitals, the spin energetics agree to within chemical
accuracy (Fig.~\ref{fig:t0_spin_energetics_error}), while the active spaces
with 8 and 9 orbitals, deviate from each other by up to $\sim 5$ kcal/mol.
Overall, as the number of the active space orbitals grows, in
Fig.~\ref{fig:t0_spin_energetics} both spin energetics show a trend towards the
spin energetics computed with the benchmark CCSD(T), as reported in Table 2 of
Ref.~\cite{Strickland2007}; $E_{\text{Q-S}} = 42.706$ kcal/mol and
$E_{\text{Q-T}} = 20.715$ kcal/mol. Moreover, the CASSCF spin energetics for
the active space (10e,12o) (not shown here) are in good agreement with the
CCSD(T) spin energetics;  $E_{\text{Q-S}} = -43.117$ kcal/mol and
$E_{\text{Q-T}} = -20.6744$ kcal/mol, respectively. This corroborates the
standard rules for picking active spaces of transition metal complexes in
Ref.~\cite{Veryazov2011}; five \ce{Fe} $3d$ orbitals plus five double-shell
\ce{Fe} $4d$ orbitals plus two ligand orbitals giving rise to $\sigma$-bonding
with \ce{Fe} $3d$ orbitals~\cite{Radon2019}.

Next, we show the same set of results for the T1 initial states.
Fig.~\ref{fig:t1_spin_energetics} shows the spin energetics for the T1 initial
states, where the triplet spin state is initialized as a multi-reference state.
Fig.~\ref{fig:t1_spin_energetics_error} shows error energy differences between
the spin energetics relative to the quintet spin state energy and
Fig.~\ref{fig:t1_relative_energies} shows energy differences between the
individual spin states. We observe for the active spaces with 5 and 7 orbitals,
the individual spin-state energies are agreement to within chemical accuracy.
The energies for the active spaces with 8 orbitals have a slight deviation
reaching up to $\sim 2$ kcal/mol while those for the active space with 9
orbitals agree to within chemical accuracy. The spin energetics follow a
similar trend (Fig.~\ref{fig:t1_spin_energetics}). For the small active spaces
with 5 to 7 orbitals, the spin energetics agree within chemical accuracy, while
the energies for 8 orbitals deviate by $\sim 4$ kcal/mol at most, and those
with 9 orbitals agree to within chemical accuracy. Similarly, as the size of
the active space grows both spin energetics show a trend towards the spin
energetics computed with CCSD(T) and CASSCF for the active space (10e,12o). The
two set of results suggest that the triplet multi-reference initial state in T1
is beneficial, increasing the accuracy of the final converged results in
comparison to the TO initial states, where all initial states are
single-references. The reported values show an improved agreement with CASSCF,
yielding energies within chemical accuracy for active spaces up to 9 orbitals
(except 8 orbitals). For both types of initial states, it is worth noting that
in some instances VQE overstabilizes spin-state energies below the
corresponding CASSCF values. The largest overstabilization observed was for
active spaces with 8 and 10 orbitals.

Due to the computational constraints, the active space for 10 orbitals is an
outlier for both sets of results. The results for this active space are
computed with $k=3$ for the $k$-UpCCGSD ansatz at 32-bit floating point
precision, hence they are expected to be less accurate than the other active
spaces computed with $k=4$ at 64-bit floating point precision. This was due to
the limits of the computational resources at our disposal (48 GB of GPU VRAM
and 502 GB of DDR4 RAM) as during the energy optimization all the $k$-UpCCGSD
cluster operators need to be in GPU memory. For instance, $k=4$ would
require keeping all 1080 sparse excitation operators in memory. See Tab.~\ref{tab:parameter_scaling}
for the number of cluster operators and parameters for each active space.

\begin{table}[!ht]
	\centering
	\begin{tabular*}{\linewidth}{@{\extracolsep{\fill}} cccc}
		\toprule
		Threshold & CAS(e,o) & $n(\{\hat{g}_i\})$ & $n(\{\theta_{i}\})$                              \\
		\midrule
		0.97      & (6e,5o)  & 240 & 220 \\
		0.95      & (8e,6o)  & 360 & 330 \\
		0.90      & (8e,7o)  & 504 & 462 \\
		0.85      & (8e,8o)  & 672 & 616 \\
		0.80      & (8e,9o)  & 864 & 792 \\
		0.70      & (8e,10o) & 990 & 1080 \\
		\bottomrule
	\end{tabular*}
	\caption{
		Comparison of the number of cluster operators ($\hat{g}$) and
		total independent variational parameters
		($\boldsymbol{\theta}$) for the $k$-UpCCGSD ansatz (fixed at
		$k=4$) across different active space sizes. For the last entry,
		when $k=3$ it has 810 cluster operators and 720 independent
		parameters.
		\label{tab:parameter_scaling}
	}
\end{table}

Despite this, the quintet spin state energy is the only value that has a
considerable deviation for the T0 initial states, and it is overstabilized by
roughly $\sim 2.5$ kcal/mol, while the singlet and triplet spin state energies
agree with the respective CASSCF energies to within $\sim 2$ kcal/mol. As a
result, the quintet-triplet and quintet-singlet relative energies from VQE and
CASSCF agree to within $\sim 5$ kcal/mol for the T0 initial states. For the T1
initial states, the singlet spin state energy has the greatest deviation,
roughly $\sim 15$ kcal/mol, while the quintet and triplet spin state energies
agree with the respective CASSCF energies to within $\sim 5$ kcal/mol. As a
result, the quintet-triplet relative energy from VQE and CASSCF agree to within
chemical accuracy, while the quintet-singlet relative energy is $\sim 12$
kcal/mol away from chemical accuracy. Overall, most of the VQE spin-state
energetics are corroborated by the CASSCF spin energetics, indicative of how
well a VQE simulation on ideal quantum hardware would perform using the
techniques outlined in this study. Further work would of course be needed to
understand the performance on current quantum hardware. See
Appendix~\ref{sec:energy_traces} for energy traces and
Appendix~\ref{sec:active_space_total_energies} for the absolute/total energies.

\subsection{Multi-reference diagnostics\label{sec:multi_reference_diagnostics}}

We now proceed to compute the multi-reference diagnostic $Z_{s(1)}$ for the
numerically computed spin-state wavefunctions in each active space. $Z_{s(1)}$
diagnoses whether a many-electron wavefunction is single- or multi-character by
estimating the degree of static correlation of the wavefunction based on the
one-orbital entropies of the individual orbitals~\cite{Stein2017}.
Wavefunctions with values of $Z_{s(1)}$ between $0.2$ and $1.0$ possess a
sufficient degree of static correlation to consider using multi-reference
methods for a qualitatively correct description, while single-reference methods
may be used reliably for wavefunctions with values that lie between $0$ and
$0.1$. See Appendix~\ref{sec:multi_reference_diagnostic_traces} for more
details.

Tab.~\ref{tab:multi_reference_diagnostics} shows the $Z_{s(1)}$ multi-reference
diagnostic for the three spin states of \ce{Fe(CH3N2)2-(OH2)} in each active
space, computed for both the T0 and T1 initial states. For all active spaces
and both types of initial states, the singlet spin states have values of
$Z_{s(1)}$ that exceed the threshold value of $0.1$, indicative of their
multi-reference character. As the size of the active space grows, these values
decrease towards the threshold value, but remain above it. In contrast, the
quintet spin states have values below the threshold value (almost zero). This
suggests that a single-reference method such as CCSD(T) can be used reliably
for the quintet spin state. Most of the triplet spin states exceed or are close
to the threshold value of $0.1$, except for (8e,10o). Apart from the active
space (8e,8o), the values in Tab.~\ref{tab:multi_reference_diagnostics} should
be taken \emph{cum grano salis}, as the $Z_{s(1)}$ multi-reference diagnostic
is less reliable for active spaces with a differing number of electrons and orbitals.
This prevents us from reaching a definite conclusion on the multi-reference
character of the spin states.

\begin{table}[!ht]
	\centering
	\begin{tabular*}{\linewidth}{@{\extracolsep{\fill}} cc ccc ccc}
		\toprule
		Threshold & CAS(e,o) & \multicolumn{3}{c}{$Z_{s(1)}$ T0} & \multicolumn{3}{c}{$Z_{s(1)}$ T1}                                     \\
		\cmidrule(lr){3-5} \cmidrule(lr){6-8}
		          &          & $S=2$                             & $S=1$                             & $S=0$  & $S=2$  & $S=1$  & $S=0$  \\
		\midrule
		0.97      & (6e,5o)  & 0.0237                            & 0.1755                            & 0.2881 & 0.0248 & 0.1757 & 0.2884 \\
		0.95      & (8e,6o)  & 0.0020                            & 0.1498                            & 0.2160 & 0.0018 & 0.1518 & 0.2367 \\
		0.90      & (8e,7o)  & 0.0217                            & 0.1251                            & 0.1750 & 0.0224 & 0.1273 & 0.1814 \\
		0.85      & (8e,8o)  & 0.0193                            & 0.1741                            & 0.1556 & 0.0193 & 0.1640 & 0.1455 \\
		0.80      & (8e,9o)  & 0.0203                            & 0.1158                            & 0.1125 & 0.0201 & 0.0989 & 0.1149 \\
		0.70      & (8e,10o) & 0.0328                            & 0.0558                            & 0.1306 & 0.0218 & 0.0709 & 0.1186 \\
		\bottomrule
	\end{tabular*}
	\caption{
		$Z_{s(1)}$ multi-reference diagnostic for different active
		spaces of \ce{Fe(CH3N2)2-(OH2)}. See
		Appendix~\ref{sec:multi_reference_diagnostic_traces} for more
		details.
		\label{tab:multi_reference_diagnostics}
	}
\end{table}

When considering the $t_{1}$ multi-reference diagnostic measured in
Ref.~\cite{Strickland2007}, the values for $t_{1}$ were found to be lie between
$0.007$ and $0.037$, suggesting that the CCSD(T) single-point energies in
Ref.~\cite{Strickland2007} may be considered to be reliable, as these values
are below the threshold of $0.05$, suggesting all the spin states are
predominantly single-reference~\cite{Jiang2012a}. However, similar to
$Z_{s(1)}$, the usefulness of $t_{1}$ in predicting the reliability of CCSD(T)
results is not always clear-cut; CCSD(T) results can match reference data of
transition metal complexes to chemical accuracy despite their $t_{1}$ values
being above the threshold value~\cite{Radon2019}.

As a further test for multi-reference character, we also compute the
orbital-pair mutual information $I_{ij}$ based on the two-orbital entropies
between all pairs of
orbitals~\cite{Rissler2006,Boguslawski2013,Boguslawski2014}. From a quantum
information theoretic sense, $I_{ij}$ quantifies the amount of information
gained about one system (orbital $i$) by observing the other system (orbital
$j$). Large values for $I_{ij}$ are indicative of interactions between orbital
$i$ and $j$, while orbital pairs that do not interact take on zero values for
$I_{i \neq j}$. Thus, $I_{ij}$ can be interpreted as a measure of orbital
interaction~\cite{Rissler2006}. Moreover, $I_{ij}$ is related to the truncation
error $\varepsilon(M)$ of fixed bond dimension $M$ matrix product states. The
performance of DMRG relies on this truncation error being small for moderately
sized bond dimensions~\cite{Ali2021}. Thus, simulating electronic wavefunctions
with large values for $I_{ij}$ would require a large bond dimension. See
Appendix~\ref{sec:orbital_interactions} for more details.

Fig.~\ref{fig:t0_orbital_interaction} and Fig.~\ref{fig:t3_orbital_interaction}
show the orbital-pair mutual information $I_{ij}$ for each active space,
computed for both the T0 and T1 initial states, respectively. The orbital-pair
mutual information $I_{ij}$ corroborates the general trend observed for the
multi-reference diagnostic $Z_{s(1)}$ in
Tab.~\ref{tab:multi_reference_diagnostics}. We find that the quintet spin
states in each active space have very small, if any, orbital-pair interactions,
while both the triplet and singlet states for both the T0 and T1 initial states
show significant interactions between orbital pairs. Similar to the $Z_{s(1)}$
multi-reference diagnostic, the singlet spin states have significant orbital
interactions in all the active spaces considered while the triplet states
follow a similar trend except for the active space (8e,10o). For this active
space, the triplet states have significantly less smaller orbital interactions
relative to other triplet states in other active spaces. For the T0 initial
states, the orbital-pair mutual information $I_{ij}$ for the (8e,10o) triplet
state closely resembles that of the quintet state (see
Fig.~\ref{fig:0.70_t0_mi}). While for T1 initial states, where the (8e,10o)
triplet state initially starts off as a multi-reference state, this
multi-reference character seems to be preserved as shown by orbital
interactions (see Fig.~\ref{fig:0.70_t3_mi}). Overall, our present results when
considering both $Z_{s(1)}$ and $I_{ij}$ seem to suggest there is some degree
of multi-reference character for both the singlet and triplet spin states in
each active space, however moderate.

\section{Conclusion\label{sec:conclusion}}

In this work, we performed numerical calculations of the spin-state energetics
of the simplified model of a deoxy-myoglobin \ce{Fe(CH3N2)2-(OH2)} for
different active spaces using the VQE algorithm, simultaneously targeting
states with different spin multiplicity (singlet, triplet and quintet). For
this, we considered single- and multi-reference initial states generated by the
$k$-UpCCGSD ansatz. We compared these spin-state energetics with those obtained
from the classical CASSCF method. Our results show good qualitative agreement
with those obtained from CASSCF, with the majority of the spin energetics and
individual spin-state energies for different active spaces within chemical
accuracy. Moreover, our results were found to tend towards the CCSD(T)
reference values reported in Ref.~\cite{Strickland2007} as the size of the
active spaces grows. A \emph{vis-\`{a}-vis} comparison between the single- and
multi-reference initial states shows that the latter approach increases the
accuracy of the final converged results. The singlet and triplet spin states
were found to be of multi-reference while the quintet spin state was found to
be close to a pure single-reference wavefunction, as evidenced by the values of
their one-orbital entropy-based multi-reference diagnostic $Z_{s(1)}$ and
orbital-pair mutual information $I_{ij}$, respectively.

An immediate direction for future work would be to improve the current
implementation of the statevector-based VQE algorithm. An area of improvement
would be implementing a method that allows us to load an arbitrary number of
cluster operators in and out of GPU memory as needed in a way that is
compatible with JAX's Just-In-Time (JIT) compilation. This would allow us to
keep only the cluster operators that are needed in GPU memory during the energy
optimization, which means one can consider ansatz with a slightly larger number
of cluster operators than those considered in this work, e.g., UCCGSD or
larger active spaces.

Another direction would be to consider the use of adaptive structure
ans\"{a}tze, such as the Adaptive Derivative-Assembled Pseudo-Trotter
ans\"{a}tze (ADAPT)~\cite{Grimsley2019} or other recent variants, e.g.,
Ref.~\cite{Ramoa_2025}. In comparison to fixed structure ans\"{a}tze such as
UCCSD and UCCGSD, adaptive ans\"{a}tze have favorable circuit depth. Moreover,
the use of adaptive ans\"{a}tze could potentially reduce the number of cluster
operators that need be kept in GPU memory during the VQE energy optimization.
This could potentially allow us to consider larger active spaces than those
considered in this work, or a similar study for the larger model of FeP(Im) in
Fig.~\ref{fig:model2_quintet}.

Alternatively, instead of a statevector simulator one can also consider the use
of tensor network-based simulators such as matrix product states (MPS),
projected entangled pair states (PEPS), multi-scale entanglement
renormalization ansatz (MERA), and related methods~\cite{Berezutskii2025}. Some
of these methods are available in JAX-based libraries like
quimb~\cite{Gray2018}. Here, variants of the adaptive structure ans\"{a}tze
such as ADAPT-VQE might be a better fit, as the resulting gate count and
circuit depth from a fixed structure ansatze like $k$-UpCCGSD ans\"{a}tze could
be computationally prohibitive. Recovering the missing contributions of dynamic
correlation using the perturbative second-order correction to the electronic
energy via multi-reference perturbation theory (MRPT) methods, as in
Ref.~\cite{Gunther2024}, could also be considered as a future direction for
this work. This would allow for a more accurate description of the spin-state
energetics of \ce{Fe(CH3N2)2-(OH2)}.

The results presented here are limited to statevector and noiseless
simulations; as a consequence, this does not fully represent the practical
considerations. Future studies should consider a shots-based simulator to mimic
the measurement process on a real quantum device. Here, one must contend with
practical considerations such as appropriately allocating the number of
measurement shots to achieve a given level of precision for the expectation
values for each energy evaluation during the optimization loop. Additionally,
future studies for noisy intermediate-scale quantum (NISQ) applications should
also incorporate realistic device noise models, such as those available in
Qiskit Aer~\cite{Qiskit2024}. Insights gained from such a study could open up a
pathway towards using near-term quantum computing hardware to perform similar
calculations for modeling the spin energetics of transition metal complexes.
The results reported here may be used as a benchmark for the performance of
VQE-based algorithms in modeling the spin energetics of transition metal
complexes. We believe that our work can help in identifying appropriate
workflows for VQE applied to model systems at this scale. Through the use of
hardware-accelerated (through GPUs, TPUs, etc.) quantum simulators and
computational methods, such as those presented here, we hope our work helps
open up a path for others to follow for simulating similarly sized strongly
correlated systems.

\section*{Acknowledgments}

The authors would like to thank Prof. Orde Munro, Prof. Yasien Sayed, Dr.
Ismail Akhalwaya, Prof. Manuel Fernandes and  Dr. Glenn Maguire for their
valuable insights and discussions at the early stages of this work. We would
also like to thank Prof. Gert Kruger for providing computational resources
where the early stages of this work were carried out. We also thank Prof.
Jeremy Harvey for providing the additional data of the simplified heme models
studied in Ref.~\cite{Strickland2007}. Finally, we thank Jane Dai for
familiarizing the authors with Schr\"{o}dinger's Jaguar software. This research was
supported by the South African National Research Foundation, the South African
Council for Scientific and Industrial Research, and the South African
Department of Science and Innovation through its Quantum Initiative program
(SAQuTI).

\bibliographystyle{apsrev4-2-title}
\bibliography{main}

\appendix
\onecolumngrid
\newpage

\counterwithin{equation}{section}
\counterwithin{table}{section}
\counterwithin{figure}{section}

\section{Molecular geometries\label{sec:molecular_geometries}}
The geometry optimization of the molecular structure of \ce{Fe(CH3N2)2-(OH2)}
from Ref.~\cite{Strickland2007} was done using Schr\"{o}dinger's Jaguar 11.5,
release 141 for the different spin states, singlet ($S = 0$), triplet ($S = 1$)
and quintet ($S = 2$). This was primarily motivated by the discrepancy of the
versions of Jaguar used in this work and Ref.~\cite{Strickland2007}. The
individual spin-state energies and relative energies are shown in
Tab.~\ref{tab:dft_abs_energies} and Tab.~\ref{tab:dft_rel_energies},
respectively. The two sets of results are in good agreement despite the use of
different versions of Jaguar. Consequently, the final B3LYP geometries for each
spin state are identical to within 3 decimal places to with those reported in
the supplementary material of Ref.~\cite{Strickland2007}.

\begin{table}[ht]
	\centering
	\begin{tabular}{lcccccc}
		\toprule
		           & \multicolumn{3}{c}{Strickland et al. 2006 \cite{Strickland2007}} & \multicolumn{3}{c}{Our work}                                                 \\
		\cmidrule(r){2-4} \cmidrule(r){5-7}
		Functional & $S=2$                                                            & $S=1$                        & $S=0$     & $S=2$     & $S=1$     & $S=0$     \\
		\midrule
		BP86       & -498.9122                                                        & -498.9107                    & -498.8746 & -498.9120 & -498.9107 & -498.8746 \\
		BLYP       & -498.6761                                                        & -498.6743                    & -498.6423 & -498.6758 & -498.6743 & -498.6423 \\
		B3PW91     & -498.7230                                                        & -498.7013                    & -498.6278 & -498.7229 & -498.7013 & -498.6628 \\
		B3P86      & -500.2039                                                        & -498.1864                    & -500.1501 & -500.1865 & -500.1865 & -500.1501 \\
		B3LYP      & -498.8420                                                        & -498.8250                    & -498.7920 & -498.8420 & -498.8250 & -498.7921 \\
		\bottomrule
	\end{tabular}
	\caption{
		Comparison of absolute DFT spin-state energies for
		\ce{Fe(CH3N2)2-(OH2)} from Ref.~\cite{Strickland2007} and after
		geometry optimization with Jaguar 11.5, release 141. The
		energies are measured in Hartrees.
		\label{tab:dft_abs_energies}
	}
\end{table}

\begin{table}[ht]
	\centering
	\begin{tabular}{lcccc}
		\toprule
		           & \multicolumn{2}{c}{Strickland et al. 2006 ~\cite{Strickland2007}} & \multicolumn{2}{c}{Our work}                                   \\
		\cmidrule(r){2-3} \cmidrule(r){4-5}
		Functional & $\Delta S = 1$                                                    & $\Delta S = 2$               & $\Delta S = 1$ & $\Delta S = 0$ \\
		\midrule
		BP86       & 0.9                                                               & 23.6                         & 0.85           & 23.48          \\
		BLYP       & 1.1                                                               & 21.2                         & 0.95           & 21.05          \\
		B3PW91     & 13.6                                                              & 37.8                         & 13.60          & 37.76          \\
		B3P86      & 10.9                                                              & 33.7                         & 10.92          & 33.69          \\
		B3LYP      & 10.7                                                              & 31.3                         & 10.74          & 31.39          \\
		\bottomrule
	\end{tabular}
	\caption{
		Comparison of relative DFT spin-state energies (relative to the
		quintet spin state energy) for \ce{Fe(CH3N2)2-(OH2)} from
		Ref.~\cite{Strickland2007} and after geometry optimization with
		Jaguar 11.5, release 141. The energies are measured in
		kcal/mol.
		\label{tab:dft_rel_energies}
	}
\end{table}
The specifications used in our calculation follow the same specifications as in Ref.~\cite{Strickland2007}, and
presented in Tab.~\ref{tab:jaguar_specs}.

\begin{table}[htbp]
	\centering
	\begin{tabular}{@{}ll@{}}
		\toprule
		\textbf{Parameter}      & \textbf{Description}                                                      \\
		\midrule
		\multicolumn{2}{l}{\textbf{Basis Set}}                                                              \\
		basis=LAC3VP,6-311G*    & LAC3VP for Fe (effective core potential) and 6-311G* for all other atoms. \\
		\midrule
		\multicolumn{2}{l}{\textbf{DFT Grid Settings}}                                                      \\
		gdftgrad=-14            & Sets the finest grid for DFT gradients.                                   \\
		gdftmed=-14             & Sets the finest grid for SCF in DFT calculations.                         \\
		gdftfine=-14            & Ensures a fine grid for DFT precision during calculations.                \\
		grid density=maximum    & Uses the highest grid density for numerical integration.                  \\
		\midrule
		\multicolumn{2}{l}{\textbf{SCF and Optimization}}                                                   \\
		maxit=5000              & Maximum allowed SCF iterations.                                           \\
		tol=1e-5                & Tolerance for SCF energy convergence.                                     \\
		rms tol=5e-6            & RMS tolerance for wavefunction convergence.                               \\
		\midrule
		\multicolumn{2}{l}{\textbf{Symmetry Handling}}                                                      \\
		idoabe=1                & Restricts symmetry to Abelian point groups.                               \\
		ipopsym=0               & Disables symmetry operations in SCF optimization.                         \\
		isymm=8                 & Allows full symmetry handling during calculations.                        \\
		\midrule
		\multicolumn{2}{l}{\textbf{Accuracy Level}}                                                         \\
		accuracy level=accurate & Ensures the most accurate calculation possible.                           \\
		\bottomrule
	\end{tabular}
	\caption{
		Specifications used in the geometry optimization of the
		molecular structure of \ce{Fe(CH3N2)2-(OH2)} with Jaguar 11.5,
		release 141.
		\label{tab:jaguar_specs}
	}
\end{table}

\section{Automated construction of active spaces\label{sec:avas_active_spaces}}

\begin{figure*}[ht!]
	\centering
	\includegraphics[width=.8\linewidth]{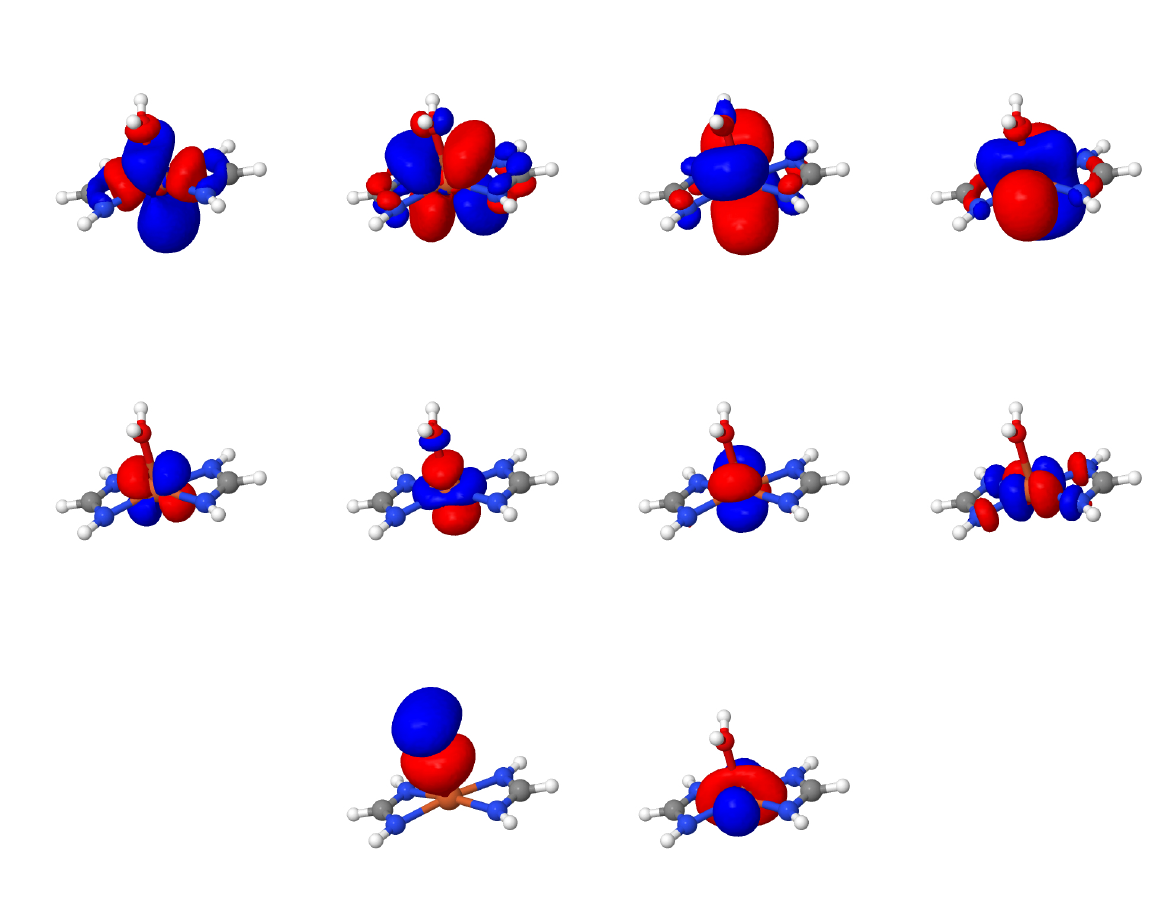}
	\caption{
		The (8e,10o) initial active space for \ce{Fe(CH3N2)2-(OH2)} as
		computed with AVAS. This active space is based on the ROHF wave
		function of the quintet ground state \ce{Fe(CH3N2)2-(OH2)} and
		is generated with a threshold value of 0.70. From top to
		bottom, the active space consists of 4 unoccupied orbitals, 4
		singly occupied orbitals, and 2 doubly occupied orbitals.
		Unlike the doubly and unoccupied orbitals, when the open-shell
		option is set to \texttt{3}, the singly occupied orbitals are
		included regardless of the threshold value. The unoccupied
		orbitals are predominantly of \ce{Fe} $4d$ character; 72\%,
		81\%, 89\% and 89\%, respectively. The first doubly occupied
		orbital has 13\% \ce{Fe} $3d$, 25\% \ce{Fe} $4d$ and 0\% \ce{O}
		$2p_{z}$, while the second one has 99 \%  \ce{Fe} $3d$, 36 \%
		\ce{Fe} 4d and 95\% \ce{O} $2p_{z}$. The surfaces depict
		isosurfaces at an isovalue of 0.025, with positive (red) and
		negative (blue) phases of the wavefunction. These isosurfaces
		were rendered using Jmol.
		\label{fig:avas_active_spaces}
	}
\end{figure*}

The active spaces for the different spin states are constructed from the B3LYP
equilibrium geometry of the quintet spin state of \ce{Fe(CH3N2)2-(OH2)} with
AVAS~\cite{Sayfutyarova2017}, as implemented in PySCF. We used the default
options in PySCF, but set the open-shell option to \texttt{3}, which ensures
that the CASCI energy always lies below the variational Hartree-Fock (HF)
energy ~\cite{Sayfutyarova2017}. We chose the active atomic orbitals \ce{Fe}
$3d$, \ce{Fe} $4d$ and \ce{O} $2p_{z}$. To include non-valence double-shell
\ce{Fe} $4d$ atom orbitals, we use the relativistic atomic natural basis set
ANO-RCC. The active space orbitals are shown in
Fig.~\ref{fig:avas_active_spaces}.

\section{Initial states\label{sec:initial_states}}

The initial states $\{\ket*{\Phi_{i}}\}$ in Eq.~\ref{eq:sa_oo_vqe_exp_val} are
constructed as occupation number vectors by specifying the spin orbital
occupations in an active space of $N$ electrons (where $N$ is even) and $M$
orbitals, where the alternating spin-up ($\alpha$) and spin-down ($\beta$)
convention is used for encoding the spin orbital occupations in qubit space via
the Jordan-Wigner mapping. Since our molecular orbitals are from a restricted
open-shell Hartree-Fock (ROHF) calculation, the aforementioned initial states
are constructed as restricted determinants~\cite{Szabo1996}, whereby the first
$K$ orbitals are doubly occupied orbitals, then the remaining orbitals are
singly occupied as far as possible. The restriction on the singly occupied
orbitals is that they all have to be $\alpha$-orbitals, \emph{i.e.} occupied by
electrons with spin-up. States constructed in this manner where all the
occupied orbitals are doubly occupied, are referred to as closed-shell
restricted determinants. When there is one or more singly occupied orbital,
they are referred to as open-shell restricted determinants. Both closed-shell
and open-shell restricted determinants are eigenfunctions of the total spin
operator squared $\hat{S}^{2}$ and $\hat{S}_{z}$, with eigenvalues
$[(n_{\alpha} - n_{\beta})^2 + 2(n_{\alpha} + n_{\beta})]/4$ and $(n_{\alpha} -
n_{\beta})/2$, where $n_{\alpha}$ and $n_{\beta}$ are the number of unpaired
$\alpha$ and $\beta$ electrons, respectively~\cite{Helgaker2014,Szabo1996}.

\begin{figure*}[!ht]
	\begin{subfigure}[t]{0.329\linewidth}
		\caption{\label{fig:t0_initial_states}}
		\includegraphics[height=0.25\textheight]{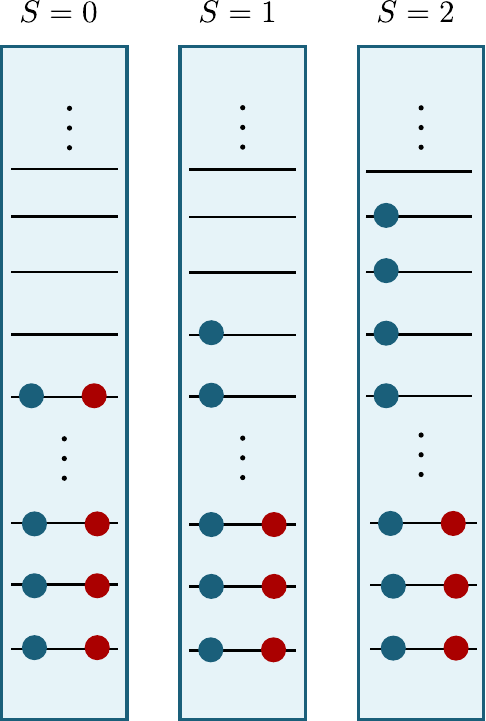}
	\end{subfigure}
	\hspace{.225\linewidth}
	\begin{subfigure}[t]{0.329\linewidth}
		\caption{\label{fig:t1_initial_states}}
		\includegraphics[height=0.25\textheight]{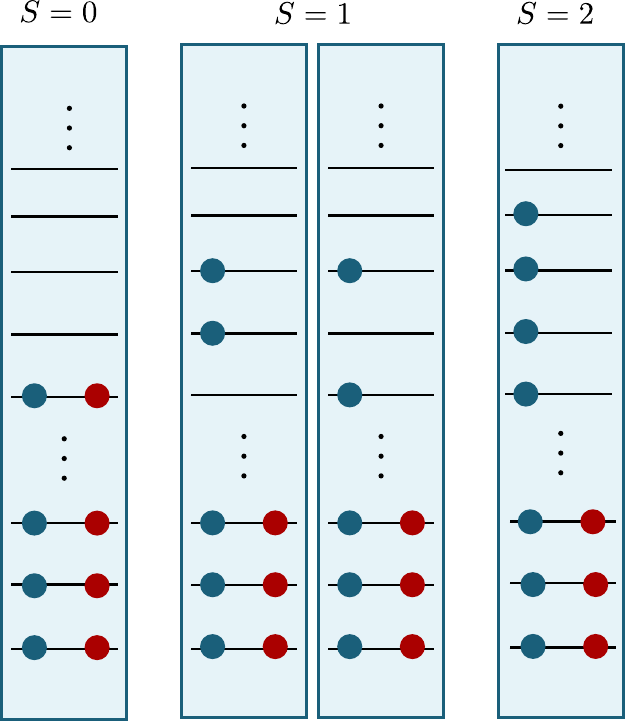}
	\end{subfigure}
	\caption{
		Schematic representation of the \textbf{(a)} T0  and \textbf{(b)} T1
		initial states for the singlet $(S = 0)$, triplet $(S = 1)$ and quintet
		$(S = 2)$ spin states, respectively. Spatial orbitals, represented by
		the horizontal bars, are doubly occupied with one $\alpha$-electron
		(blue) and one $\beta$-electron (red) as far as possible, then the
		orbitals are singly occupied by $\alpha$-electrons (blue).
		\label{fig:initial_states}
	}
\end{figure*}

Following the above procedure, the T0 initial states are constructed as
follows. For the singlet spin state $\ket*{\Phi_{0}}$, the first $N/2$ orbitals
are doubly occupied. Since there is an equal number of $\alpha$- and
$\beta$-electrons, \emph{i.e.} no unpaired electrons, the resulting determinant
forms a state with a total intrinsic spin of $S = -1/2 + 1/2 + \cdots -1/2 +
	1/2 = 0$ and $\ev*{\hat{S}^2} = [(0 - 0)^2 + 2(0 + 0)]/4 = 0$. For the triplet
spin state $\ket*{\Phi_{1}}$, the first $(N-2)/2$ orbitals are doubly occupied, then
the orbitals $(N-2)/2 + 1$ to $(N-2)/2 + 2$ are singly occupied by $\alpha$-electrons. The
determinant therefore forms a state with two unpaired $\alpha$-electrons and no
unpaired $\beta$-electrons, that has a total intrinsic spin of $S = 2 \times
	1/2 = 1$ and $\ev*{\hat{S}^2} = [(2 - 0)^2 + 2(2 + 0)]/4 = 2$. Similarly, for
the quintet spin state $\ket*{\Phi_{2}}$, the first $(N-4)/2$ orbitals are doubly
occupied then the orbitals $(N-4)/2 + 1$ to $(N-4)/2 + 4$ are singly occupied with
$\alpha$-electrons such that there are four unpaired $\alpha$-electrons and no
unpaired $\beta$-electrons, giving a state with a total intrinsic spin of $S =
	4 \times 1/2 = 2$ and $\ev*{\hat{S}^2} = [(4 - 0)^2 + 2(4 + 0)]/4 = 6$. See
Fig.~\ref{fig:t0_initial_states} for a schematic representation of the
described states. For instance, for an active space of 6 electrons and 5
orbitals, the initial states would take the following form in qubit space:

\begin{align}
	\ket*{\Phi_{0}} & = \ket*{1111110000} \nonumber \\
	\ket*{\Phi_{1}} & = \ket*{1111101000} \nonumber \\
	\ket*{\Phi_{2}} & = \ket*{1110101010},
\end{align}
where we used the alternating spin-up ($\alpha$) and spin-down ($\beta$)
convention for encoding the spin orbital occupations. For the T1 initial
states, the preparation of singlet and quintet spin states remains unchanged,
but the triplet spin state is prepared as a uniform superposition of two
occupation number vectors. The first occupation number vector has the first
$(N-2)/2$ orbitals doubly occupied then the orbitals $(N-2)/2 + 2$ to $(N-2)/2
+ 3$ are singly occupied in $\alpha$-orbitals, skipping orbital $(N-2)/2 + 1$.
For the other occupation number vector, the first $(N-2)/2$ orbitals are doubly
occupied then the orbitals $(N-2)/2 + 1$ to $(N-2)/2 + 3$ are singly occupied,
skipping orbital $(N-2)/2 + 2$, see Fig.~\ref{fig:t1_initial_states}.

\noindent
Since both states have no unpaired $\beta$-electrons and two unpaired
$\alpha$-electrons, the resulting state is a triplet state with a total
intrinsic spin of $S = 2 \times 1/2 = 1$ and $\ev*{\hat{S}^2} = [(2 - 0)^2 +
	2(2 + 0)]/4 = 2$. For an active space of 6 electrons and 5 orbitals, the
multi-reference triplet state would take the following form in qubit space:

\begin{align}
	\ket*{\Phi_{1}} & = \frac{1}{\sqrt{2}}(\ket*{1111001010}  - \ket*{1111100010}).
\end{align}
The uniform superposition of these two occupation number vectors forms a state
that is also a triplet state since both occupation number vectors are triplet
states.

\section{Energy traces\label{sec:energy_traces}}
\begin{figure}[ht!]
	\begin{subfigure}[t]{0.329\linewidth}
		\caption{}
		\includegraphics[width=\linewidth]{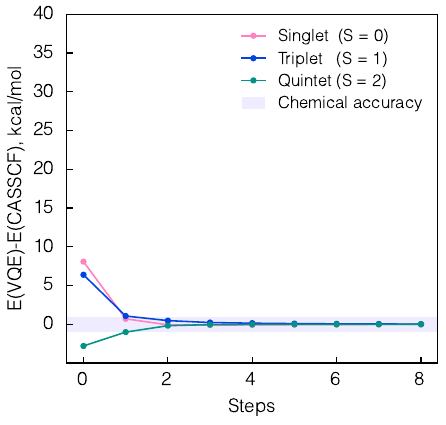}
	\end{subfigure}
	\begin{subfigure}[t]{0.329\linewidth}
		\caption{}
		\includegraphics[width=\linewidth]{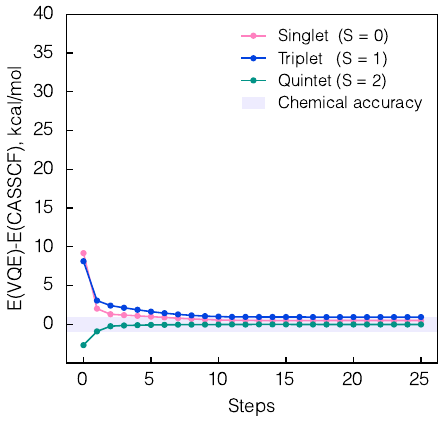}
	\end{subfigure}
	\begin{subfigure}[t]{0.329\linewidth}
		\caption{}
		\includegraphics[width=\linewidth]{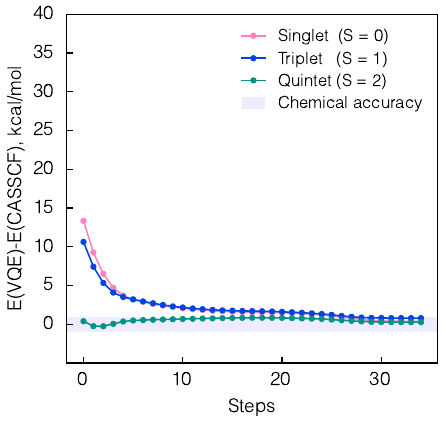}
	\end{subfigure}
	\begin{subfigure}[t]{0.329\linewidth}
		\caption{}
		\includegraphics[width=\linewidth]{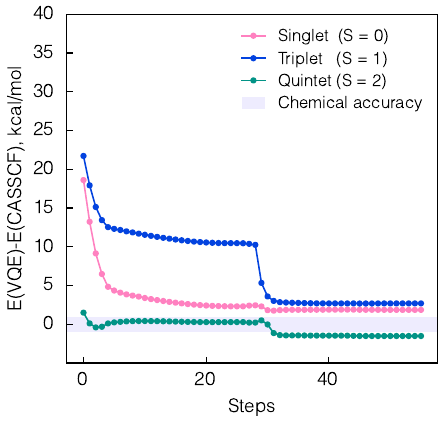}
	\end{subfigure}
	\begin{subfigure}[t]{0.329\linewidth}
		\caption{}
		\includegraphics[width=\linewidth]{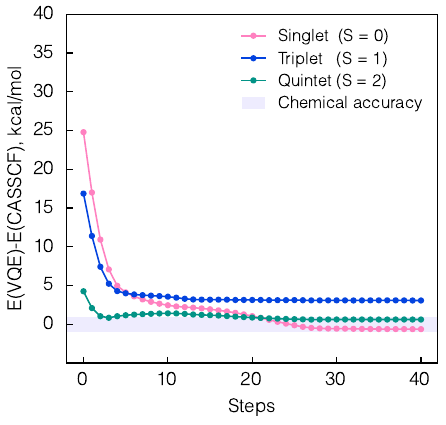}
	\end{subfigure}
	\begin{subfigure}[t]{0.329\linewidth}
		\caption{}
		\includegraphics[width=\linewidth]{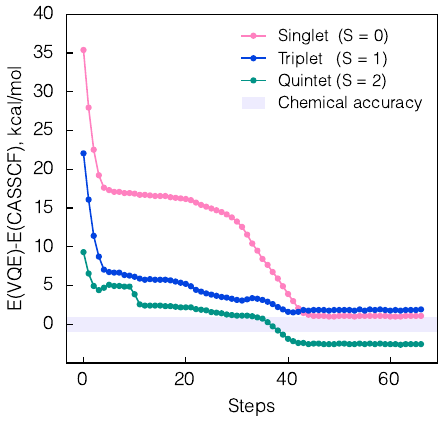}
	\end{subfigure}
	\caption{
		Energy traces for the singlet $(S = 0)$, triplet $(S = 1)$ and
		quintet $(S = 2)$ spin states for the T0 initial states,
		respectively. The plots show relative energies (with respective
		to CASSCF energies) during VQE optimization for the active
		spaces \textbf{(a)} (6e,5o), \textbf{(b)} (8e,6o), \textbf{(c)}
		(8e,7o), \textbf{(d)} (8e,8o), \textbf{(e)} (8e,9o) and
		\textbf{(f)} (8e,10o).
		\label{fig:t0_energy_traces}
	}
\end{figure}

\begin{figure*}[ht!]
	\begin{subfigure}[t]{0.329\linewidth}
		\caption{}
		\includegraphics[width=\linewidth]{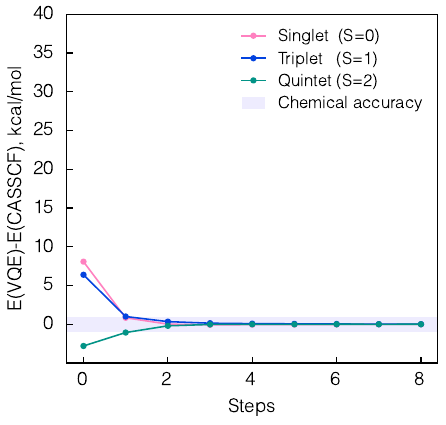}
	\end{subfigure}
	\begin{subfigure}[t]{0.329\linewidth}
		\caption{}
		\includegraphics[width=\linewidth]{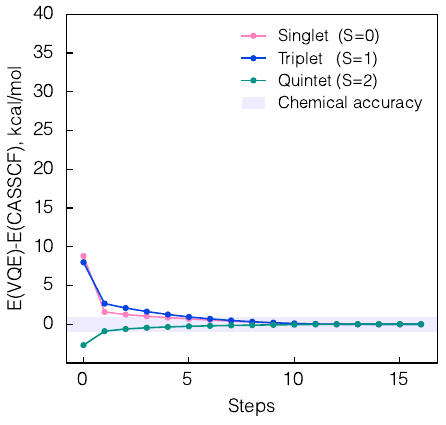}
	\end{subfigure}
	\begin{subfigure}[t]{0.329\linewidth}
		\caption{}
		\includegraphics[width=\linewidth]{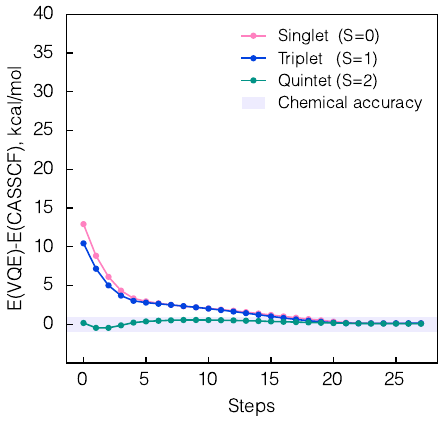}
	\end{subfigure}
	\begin{subfigure}[t]{0.329\linewidth}
		\caption{}
		\includegraphics[width=\linewidth]{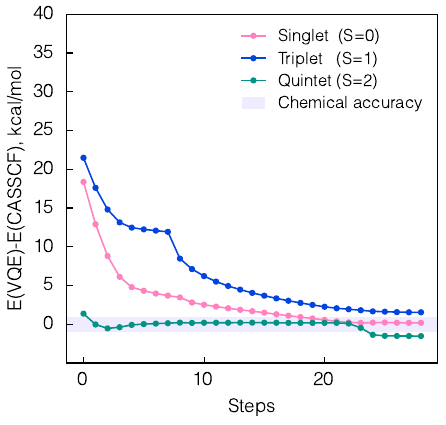}
	\end{subfigure}
	\begin{subfigure}[t]{0.329\linewidth}
		\caption{}
		\includegraphics[width=\linewidth]{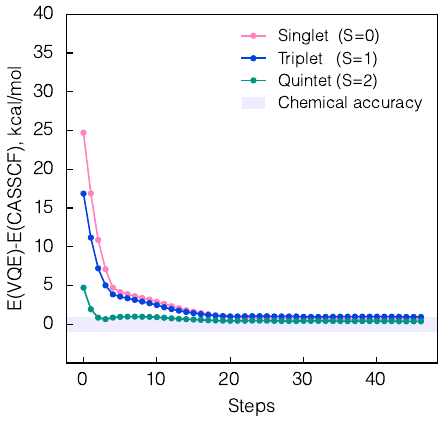}
	\end{subfigure}
	\begin{subfigure}[t]{0.329\linewidth}
		\caption{}
		\includegraphics[width=\linewidth]{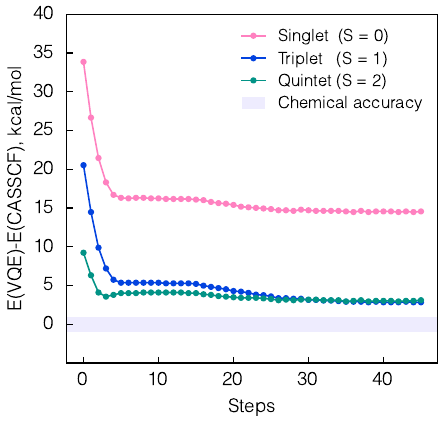}
	\end{subfigure}
	\caption{
		Energy traces for the singlet $(S = 0)$, triplet $(S = 1)$ and
		quintet $(S = 2)$ spin states for the T1 initial states,
		respectively. The plots show relative energies (with respective
		to CASSCF energies) during VQE optimization for the active
		spaces \textbf{(a)} (6e,5o), \textbf{(b)} (8e,6o), \textbf{(c)}
		(8e,7o), \textbf{(d)} (8e,8o), \textbf{(e)} (8e,9o) and
		\textbf{(f)} (8e,10o).
		\label{fig:t1_energy_traces}
	}
\end{figure*}

All the VQE energies reported in this work were obtained with 100 or fewer
macro cycles, where each macro cycle has one micro cycle. The convergence
tolerance for the VQE optimization was set to $10^{-7}$ Hartrees. The VQE
energy traces for the T0 and T1 initial states are shown in
Fig.~\ref{fig:t0_energy_traces} and Fig.~\ref{fig:t1_energy_traces},
respectively. The energy traces show the convergence of the VQE optimization
for the different active spaces.

\newpage

\section{Active space total electronic energies \label{sec:active_space_total_energies}}
All the state-averaged CASSCF total energies computed with PySCF were obtained
using 100 or fewer macro cycles, where each macro cycle has 4 micro cycles. We
also set the internal rotation flag to true to enable orbital optimization, and
the convergence tolerance was set to $10^{-7}$ Hartrees. The CASSCF total
energies for different active spaces are shown in
Tab.~\ref{tab:casscf_total_energies}. The active spaces were also studied with
the state-averaged DMRG active solver available, BLOCK2~\cite{Zhai2021}. We
follow the same specification as in the main text for other simulations. In all
the DMRG calculations, we use the maximum bond dimension $M = 10000$. This
value specifies the dimensions for the matrix product states (MPS) expansion in
the DMRG wavefunction. Tab.~\ref{tab:dmrg_total_energies} shows the DMRGSCF
total energies for the different active spaces. Some of the DMRG energies do
not numerically convergence to the corresponding CASSCF energies. In
particular, all the triplet states show a deviation of at least 6 kcal/mol and
at most 16 kcal/mol. This is a situation that can be possibly improved if we
use an even larger value for the maximum bond dimension $M$, and/or, the
inclusion of a subsequent calculation to account for dynamic correlation, as
done in Ref.~\cite{Freitag2017} for Model 2. Finally, for completeness sake, we
also show the corresponding VQE total energies for the T0 and T1 initial states
in Tab.~\ref{tab:vqe_t0_total_energies} and
Tab.~\ref{tab:vqe_t1_total_energies}, respectively.

\begin{table}[!ht]
	\centering
	\begin{tabular}{ccccc}
		\toprule
		Threshold & CAS(e,o) & $S=2$ & $S=1$ & $S=0$ \\
		\midrule
		0.97      & (6e,5o)  & -1635.61685958192 & -1635.55436899236  & -1635.51452926361 \\
		0.95      & (8e,6o)  & -1635.61702670230 & -1635.55704298083  & -1635.51595704288 \\
		0.90      & (8e,7o)  & -1635.62994662028 & -1635.56811952052  & -1635.52715816816 \\
		0.85      & (8e,8o)  & -1635.63983300113 & -1635.59418318484  & -1635.55428133181 \\
		0.80      & (8e,9o)  & -1635.65263705873 & -1635.60967232162  & -1635.58842121796 \\
		0.70      & (8e,10o) & -1635.67181235922 & -1635.63048937366  & -1635.61132989370 \\
		\bottomrule
	\end{tabular}
	\caption{
	   State-averaged CASSCF total energies (in Hartrees) for varying active
	   spaces for \ce{Fe(CH3N2)2-(OH2)}. The calculations were performed
	   with PySCF's CASSCF active solver where the orbital optimization
	   enabled and a convergence tolerance of $10^{-7}$ Hartrees. Results are
	   shown for spin states $S=2$ (quintet), $S=1$ (triplet), and $S=0$
	   (singlet).
	   \label{tab:casscf_total_energies}
	}
\end{table}

\begin{table}[!ht]
	\centering
	\begin{tabular}{ccccc}
		\toprule
		Threshold & CAS(e,o) & $S=2$ & $S=1$ & $S=0$ \\
		\midrule
		0.97      & (6e,5o)  & -1635.61748771604 (-0.3942) & -1635.54336503204 (6.9051)  & -1635.51440768278 (0.7163) \\
		0.95      & (8e,6o)  & -1635.61748578793 (-0.2880) & -1635.54668368805 (6.5005)  & -1635.51440768278 (0.9722) \\
		0.90      & (8e,7o)  & -1635.62932368639 (0.3909)  & -1635.55658152886 (7.2401)  & -1635.52403182665 (1.9618) \\
		0.85      & (8e,8o)  & -1635.64043923284 (-0.3804) & -1635.56767937498 (16.6314) & -1635.55010006310 (2.6239) \\
		0.80      & (8e,9o)  & -1635.65143689001 (0.7531)  & -1635.58689641496 (14.2921) & -1635.58904810602 (-0.3934) \\
		0.70      & (8e,10o) & -1635.67391658396 (-1.3204) & -1635.61151635941 (11.9057) & -1635.60487687452 (4.0493) \\
		\bottomrule
	\end{tabular}
	\caption{
		State-averaged DMRGSCF total energies (in Hartrees) computed
		using the BLOCK2 solver. A maximum bond dimension of $M =
		10000$ was used for the MPS expansion. Similar to the CASSCF
		total energies, we enabled orbital optimization and set the
		convergence tolerance to $10^{-7}$ Hartrees. The values in the
		parenthesis show the difference between DMRGSCF and CASSCF
		total energies in kcal/mol. \label{tab:dmrg_total_energies}
	}
\end{table}

\begin{table}[!h]
	\centering
	\begin{tabular}{ccccc}
		\toprule
		Threshold & CAS(e,o) & $S=2$ & $S=1$ & $S=0$ \\
		\midrule
		0.97      & (6e,5o)  & -1635.61686326614 (-0.0023) & -1635.55433622073 (0.0206)  & -1635.51455485201 (-0.0161)  \\
		0.95      & (8e,6o)  & -1635.61707310960 (-0.0291) & -1635.55556213798 (0.9292) & -1635.51519049652 (0.4810) \\
		0.90      & (8e,7o)  & -1635.62858437997 (0.8548) & -1635.56537603238 (1.7216) & -1635.52467472480 (1.5584) \\
		0.85      & (8e,8o)  & -1635.64216109510 (-1.4609) & -1635.58987726650 (2.7020)  & -1635.55129548117 (1.8736) \\
		0.80      & (8e,9o)  & -1635.65166508504 (0.6099) & -1635.60475233426 (3.0873) & -1635.58940281813 (-0.6160) \\
		0.70$^{\dagger}$     & (8e,10o) & -1635.67578125000 (-2.4905) & -1635.62744140625 (1.9126)  & -1635.60961914062 (1.0735) \\
		\bottomrule
	\end{tabular}
	\caption{
		State-averaged VQE total energies (in Hartrees) for the T0
		initial states. Similar to the CASSCF total energies, we
		enabled orbital optimization and set the convergence tolerance
		to $10^{-7}$ Hartrees. The values in the parenthesis show the
		difference between VQE and CASSCF total energies in kcal/mol.
		\label{tab:vqe_t0_total_energies}
	}
\end{table}

\begin{table}[!h]
	\centering
	\begin{tabular}{ccccc}
		\toprule
		Threshold & CAS(e,o) & $S=2$ & $S=1$ & $S=0$ \\
		\midrule
		0.97      & (6e,5o)  & -1635.61686365951 (-0.0026) & -1635.55434895992 (0.0126)  & -1635.51454460721 (-0.0096) \\
		0.95      & (8e,6o)  & -1635.61702691132 (-0.0001) & -1635.55704322391 (-0.0002) & -1635.51595656683 (0.0003) \\
		0.90      & (8e,7o)  & -1635.62951913152 (0.2683)  & -1635.56684842619 (0.7976)  & -1635.52656299587 (0.3735) \\
		0.85      & (8e,8o)  & -1635.64226277691 (-1.5247) & -1635.59170490247 (1.5551)  & -1635.55400530182 (0.1732) \\
		0.80      & (8e,9o)  & -1635.65203358914 (0.3787)  & -1635.60817201274 (0.9415)  & -1635.58731307311 (0.6954) \\
		0.70$^{\dagger}$      & (8e,10o) & -1635.66699218750 (3.0247) & -1635.62585449218 (2.9084) & -1635.58825683593 (14.4786) \\
		\bottomrule
	\end{tabular}
	\caption{
		State-averaged VQE total energies (in Hartrees) for the initial
		state T1. Similar to the CASSCF total energies, we enabled
		orbital optimization and set the convergence
		tolerance to $10^{-7}$ Hartrees. The values in the parenthesis
		show the difference between VQE and CASSCF total energies in
		kcal/mol.
		\label{tab:vqe_t1_total_energies}
	}
\end{table}

\newpage

\section{Multi-reference diagnostic traces\label{sec:multi_reference_diagnostic_traces}}
\begin{figure*}[ht!]
	\begin{subfigure}[t]{0.329\linewidth}
		\caption{}
		\includegraphics[width=\linewidth]{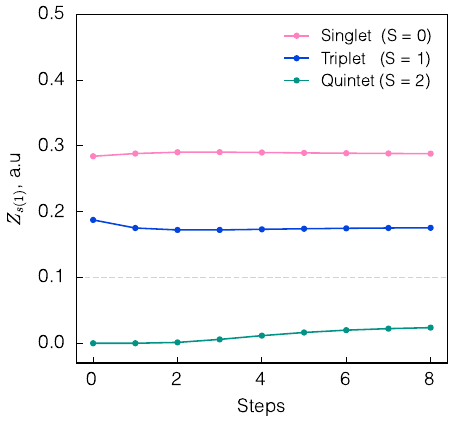}
	\end{subfigure}
	\begin{subfigure}[t]{0.329\linewidth}
		\caption{}
		\includegraphics[width=\linewidth]{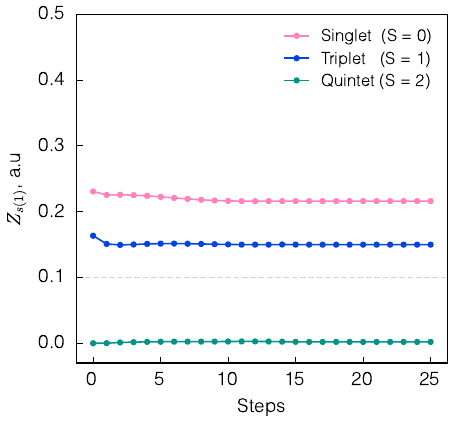}
	\end{subfigure}
	\begin{subfigure}[t]{0.329\linewidth}
		\caption{}
		\includegraphics[width=\linewidth]{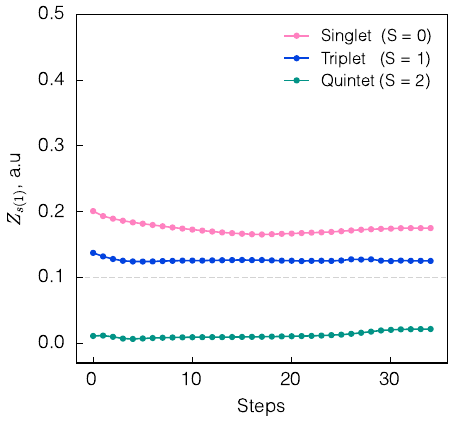}
	\end{subfigure}
	\begin{subfigure}[t]{0.329\linewidth}
		\caption{}
		\includegraphics[width=\linewidth]{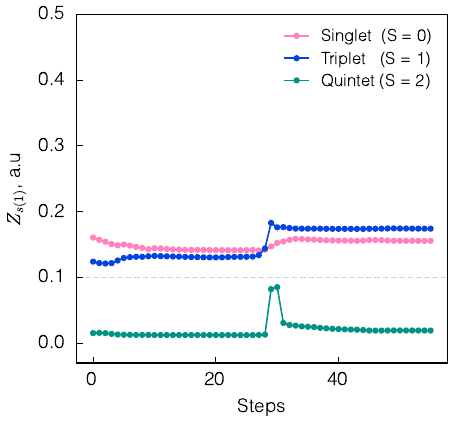}
	\end{subfigure}
	\begin{subfigure}[t]{0.329\linewidth}
		\caption{}
		\includegraphics[width=\linewidth]{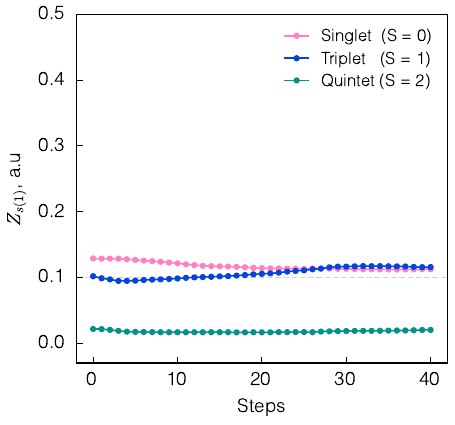}
	\end{subfigure}
	\begin{subfigure}[t]{0.329\linewidth}
		\caption{}
		\includegraphics[width=\linewidth]{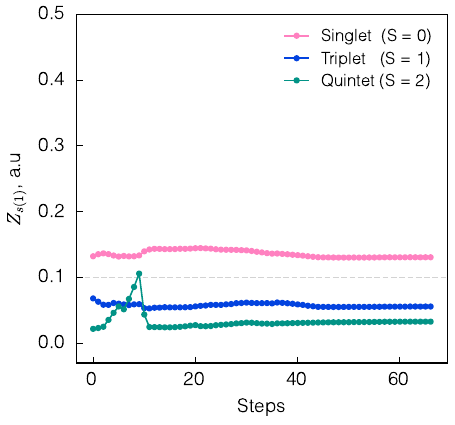}
	\end{subfigure}
	\caption{
		$Z_{s(1)}$ multi-reference diagnostic traces for the singlet
		$(S = 0)$, triplet $(S = 1)$ and quintet $(S = 2)$ spin states
		for the T0 initial states, respectively. The plots show the
		values of $Z_{s(1)}$ during VQE optimization for the active
		spaces \textbf{(a)} (6e,5o), \textbf{(b)} (8e,6o), \textbf{(c)}
		(8e,7o), \textbf{(d)} (8e,8o), \textbf{(e)} (8e,9o) and
		\textbf{(f)} (8e,10o).
		\label{fig:t0_Z_s(1)}
	}
\end{figure*}

\begin{figure*}[ht!]
	\begin{subfigure}[t]{0.329\linewidth}
		\caption{}
		\includegraphics[width=\linewidth]{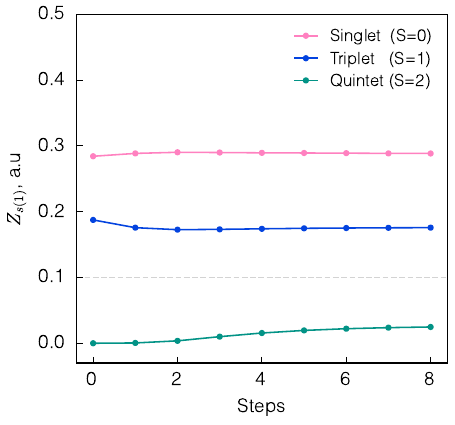}
	\end{subfigure}
	\begin{subfigure}[t]{0.329\linewidth}
		\caption{}
		\includegraphics[width=\linewidth]{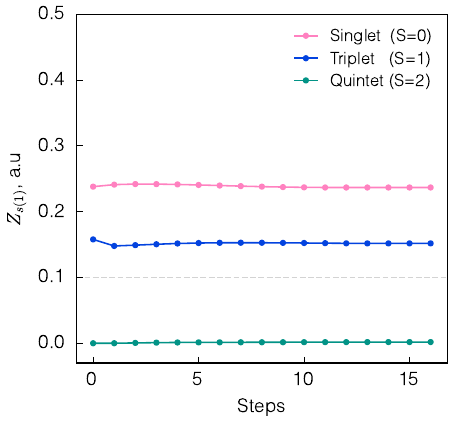}
	\end{subfigure}
	\begin{subfigure}[t]{0.329\linewidth}
		\caption{}
		\includegraphics[width=\linewidth]{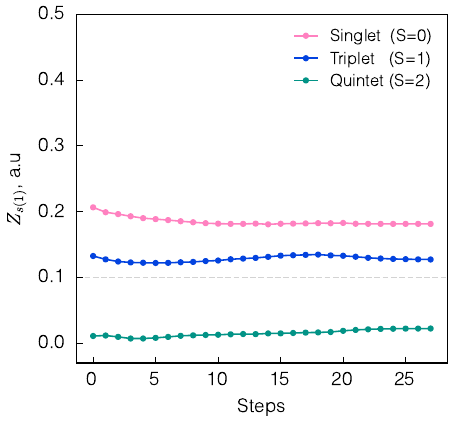}
	\end{subfigure}
	\begin{subfigure}[t]{0.329\linewidth}
		\caption{}
		\includegraphics[width=\linewidth]{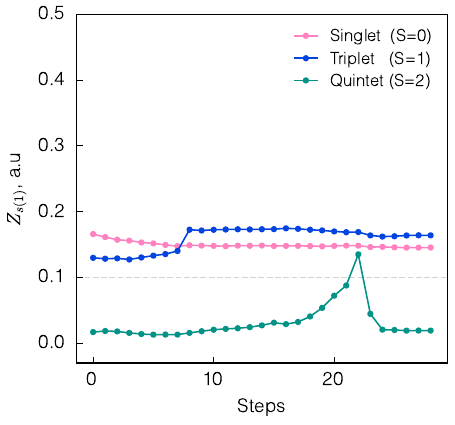}
	\end{subfigure}
	\begin{subfigure}[t]{0.329\linewidth}
		\caption{}
		\includegraphics[width=\linewidth]{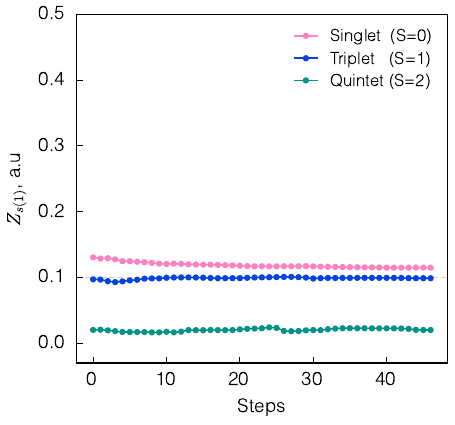}
	\end{subfigure}
	\begin{subfigure}[t]{0.329\linewidth}
		\caption{}
		\includegraphics[width=\linewidth]{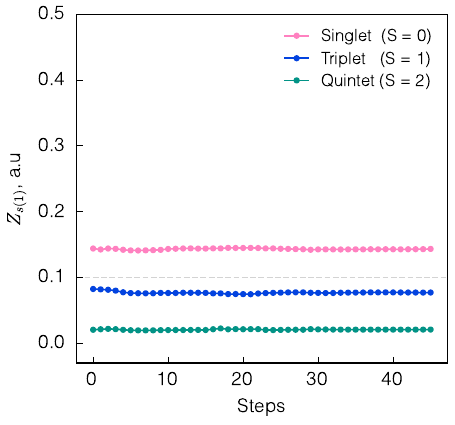}
	\end{subfigure}
	\caption{
		$Z_{s(1)}$ multi-reference diagnostic traces for the singlet
		$(S = 0)$, triplet $(S = 1)$ and quintet $(S = 2)$ spin states
		for the T1 initial states, respectively. The plots show the
		values of $Z_{s(1)}$ during VQE optimization for the active
		spaces \textbf{(a)} (6e,5o), \textbf{(b)} (8e,6o), \textbf{(c)}
		(8e,7o), \textbf{(d)} (8e,8o), \textbf{(e)} (8e,9o) and
		\textbf{(f)} (8e,10o).
		\label{fig:t1_Z_s(1)}
	}
\end{figure*}

For an active space with $L$ orbitals, the $Z_{s(1)}$ multi-reference diagnostic is given by ~\cite{Stein2017}

\begin{align}
	Z_{s(1)} = \frac{1}{L\ln{4}}\sum_{i}^{L} s_i(1).
\end{align}
Here, $s_{i}(1)$ is the one-orbital entropy of orbital $i$ given by
\begin{align}
	s_{i}(1) = -\sum_{\alpha}^{4} \omega_{\alpha, i}\ln{\omega_{\alpha, i}},
	\label{eq:one_orbital_entropy}
\end{align}
where $\omega_{\alpha, i}$ is the eigenvalue of the one-orbital reduced density
matrix for orbital $i$. The eigenvalues of the one-orbital reduced density matrix
for orbital $i$ are given by

\begin{align}
	\{\omega_{\alpha, i}\} = \{
	1 - \gamma^{i}_{i} - \gamma^{\bar{i}}_{\bar{i}} + \Gamma^{i\bar{i}}_{i\bar{i}},\>
	\gamma^{i}_{i} - \Gamma^{i\bar{i}}_{i\bar{i}},\>
	\gamma^{\bar{i}}_{\bar{i}} - \Gamma^{i\bar{i}}_{i\bar{i}},\>
	\Gamma^{i\bar{i}}_{i\bar{i}}
	\},
\end{align}
where unbarred and barred indices denote $\alpha$- and $\beta$-electrons, and
$\gamma^{i}_{j} = \ev*{\hat{a}^{\dagger}_{i}\hat{a}_{j}}$ and $\Gamma^{ij}_{kl}
= \ev*{\hat{a}^{\dagger}_{i}\hat{a}^{\dagger}_{j}\hat{a}_{l}\hat{a}_{k}}$ are
the spin-independent one- and two-particle reduced density matrices,
respectively. In our work, the one- and two-particle reduced density matrices
are computed with respect to the state vectors at the end of a VQE optimization
cycle. Fig.~\ref{fig:t0_Z_s(1)} and Fig.~\ref{fig:t1_Z_s(1)} show the
$Z_{s(1)}$ multi-reference diagnostic traces for the T0 and T1 initial states,
respectively.

\clearpage

\section{Orbital-pair mutual information \label{sec:orbital_interactions}}
\begin{figure*}[ht!]
	\centering
	\begin{subfigure}[t]{0.16\textwidth}
		\caption{}
		\includegraphics[width=\textwidth]{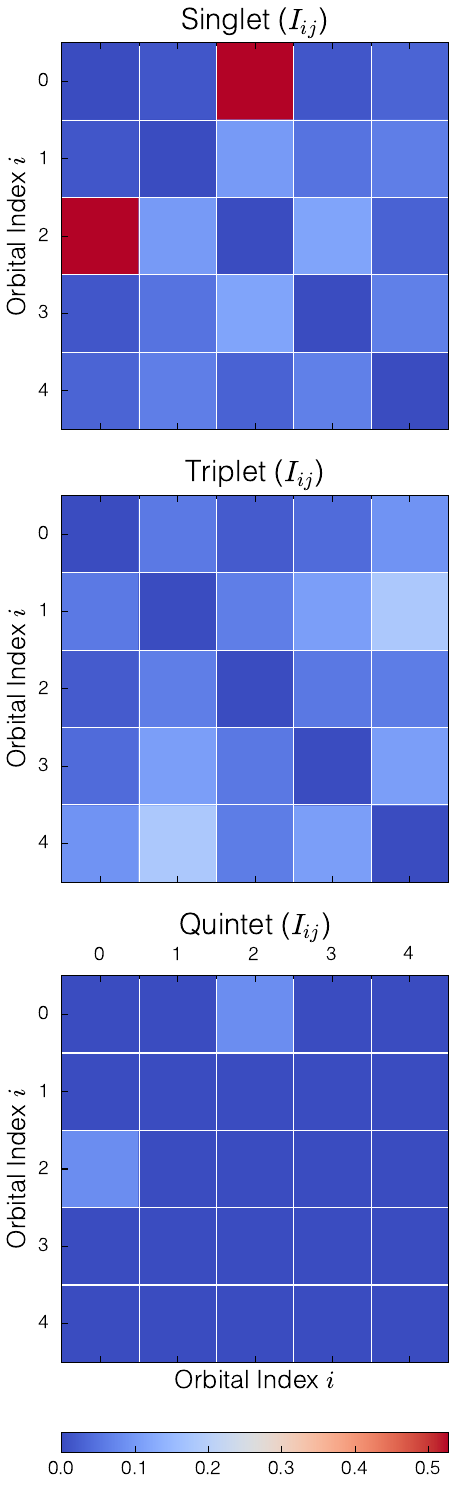}
		\label{fig:0.97_t0_mi}
	\end{subfigure}
	\begin{subfigure}[t]{0.16\textwidth}
		\caption{}
		\includegraphics[width=\textwidth]{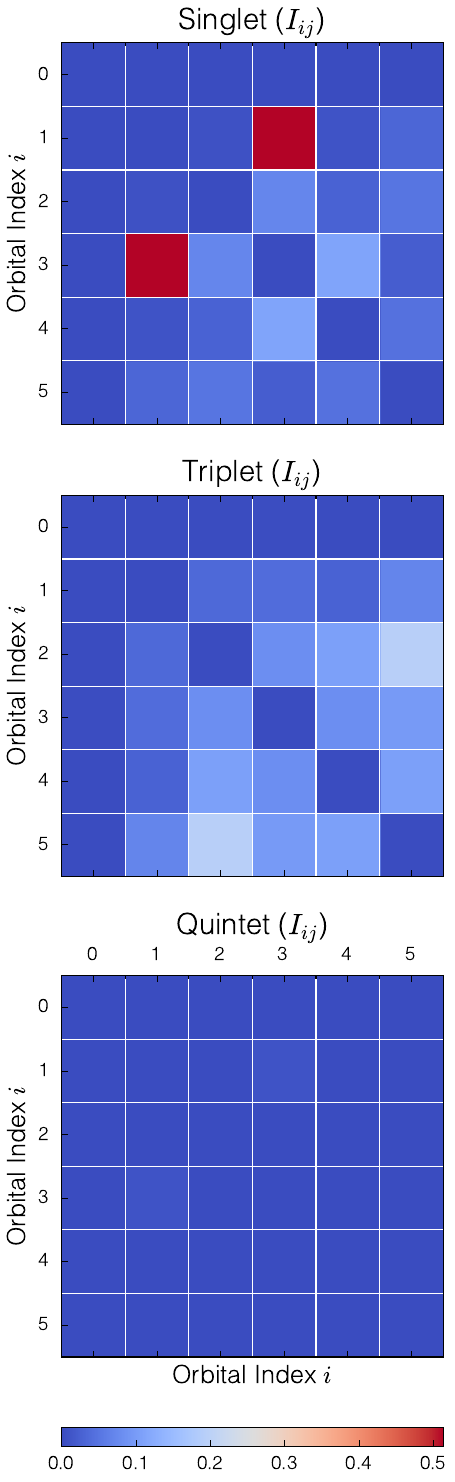}
		\label{fig:0.95_t0_mi}
	\end{subfigure}
	\begin{subfigure}[t]{0.16\textwidth}
		\caption{}
		\includegraphics[width=\textwidth]{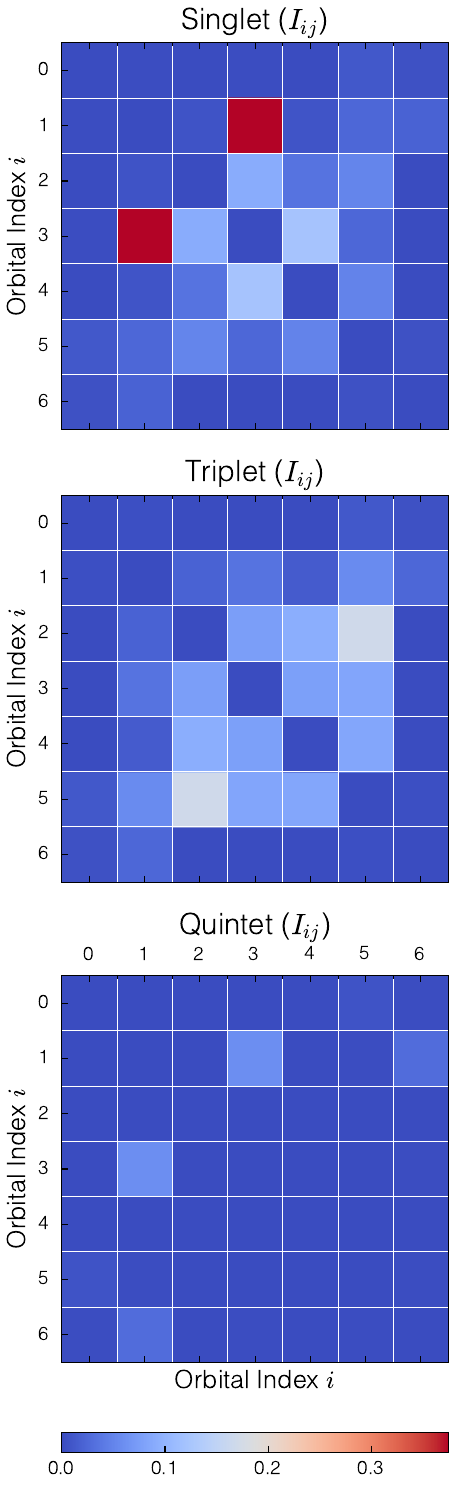}
		\label{fig:0.90_t0_mi}
	\end{subfigure}
	\begin{subfigure}[t]{0.16\textwidth}
		\caption{}
		\includegraphics[width=\textwidth]{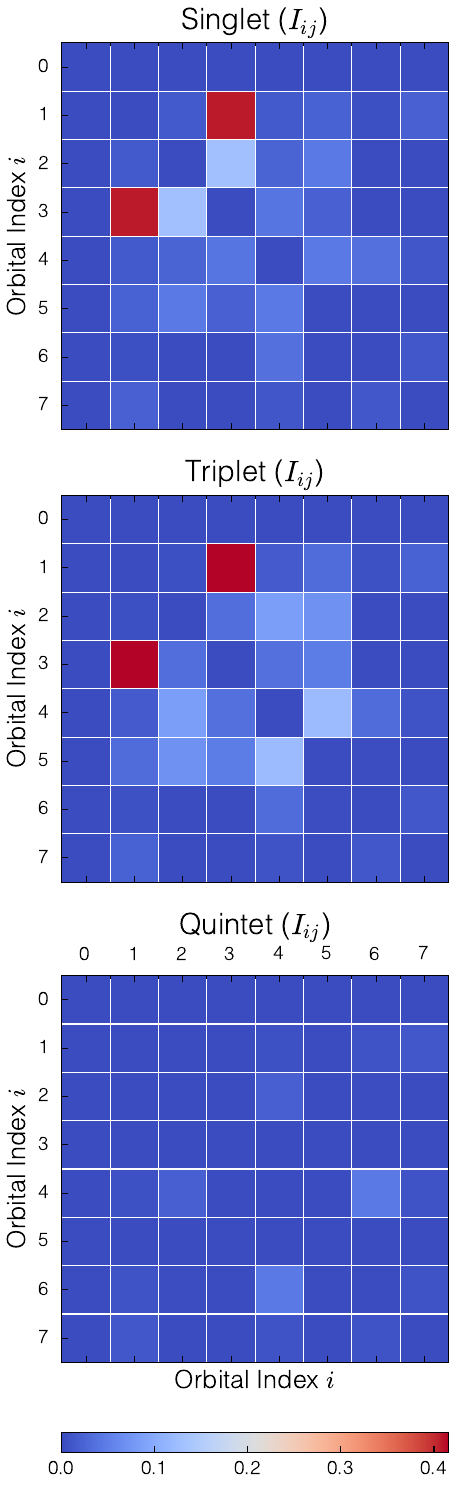}
		\label{fig:0.85_t0_mi}
	\end{subfigure}
	\begin{subfigure}[t]{0.16\textwidth}
		\caption{}
		\includegraphics[width=\textwidth]{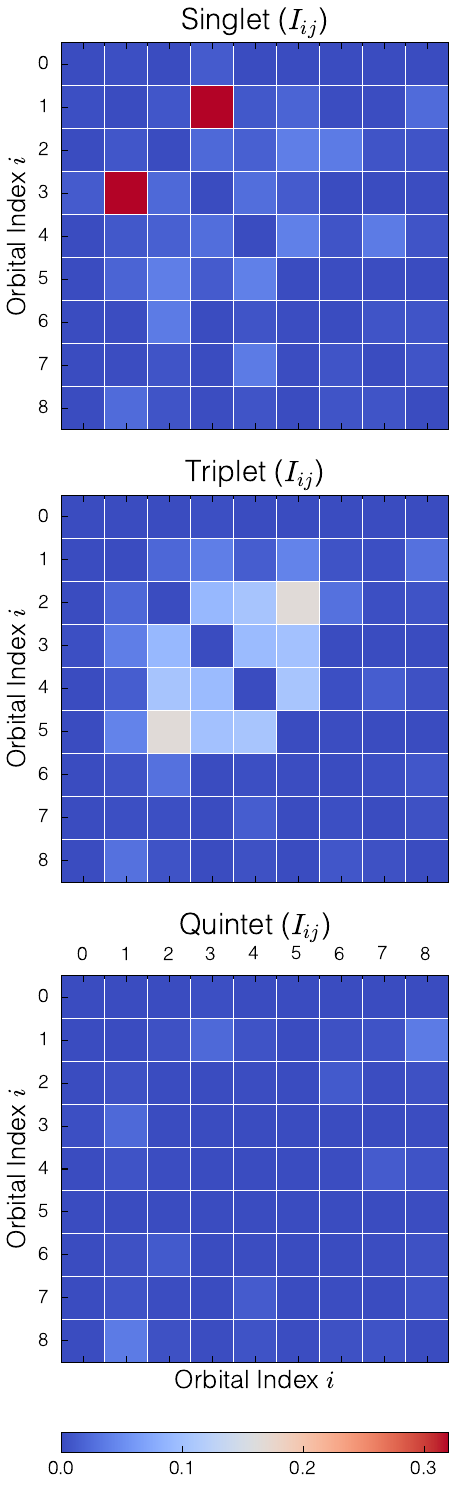}
		\label{fig:0.80_t0_mi}
	\end{subfigure}
	\begin{subfigure}[t]{0.16\textwidth}
		\caption{}
		\includegraphics[width=\textwidth]{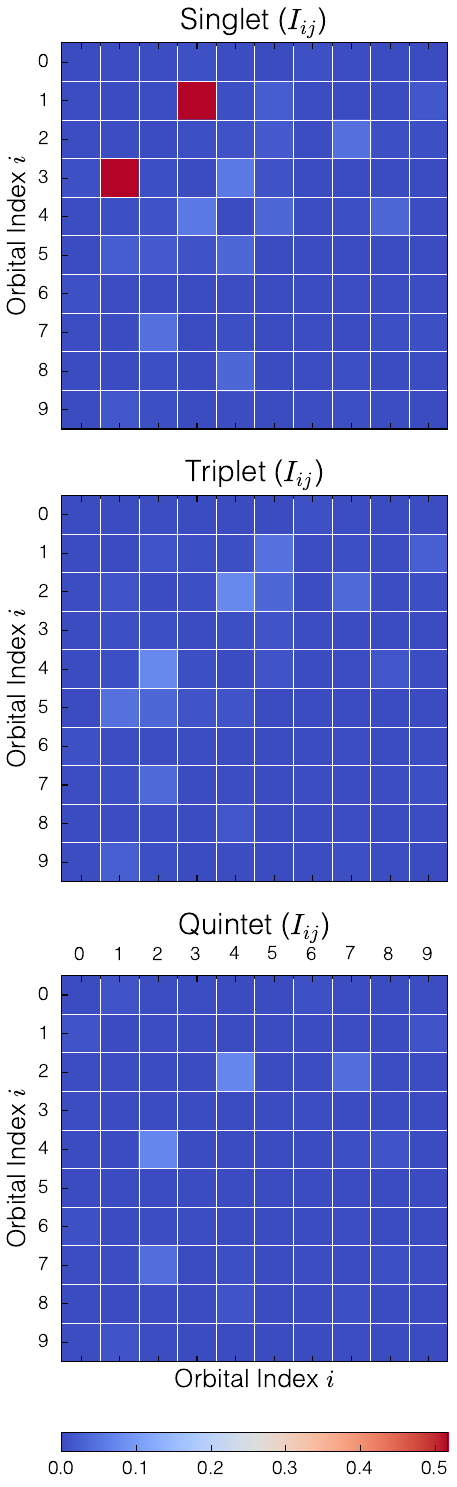}
		\label{fig:0.70_t0_mi}
	\end{subfigure}
	\caption{
		Orbital interactions for the singlet $(S = 0)$, triplet $(S =
		1)$ and quintet $(S = 2)$ spin states for T0 initial states,
		respectively. The matrix plots show $I_{ij}$ for the final and
		converged spin states for the active spaces \textbf{(a)}
		(6e,5o), \textbf{(b)} (8e,6o), \textbf{(c)} (8e,7o),
		\textbf{(d)} (8e,8o), \textbf{(e)} (8e,9o) and \textbf{(f)}
		(8e,10o).
		\label{fig:t0_orbital_interaction}
	}
\end{figure*}

\begin{figure*}[ht!]
	\centering
	\begin{subfigure}[t]{0.16\textwidth}
		\caption{}
		\includegraphics[width=\textwidth]{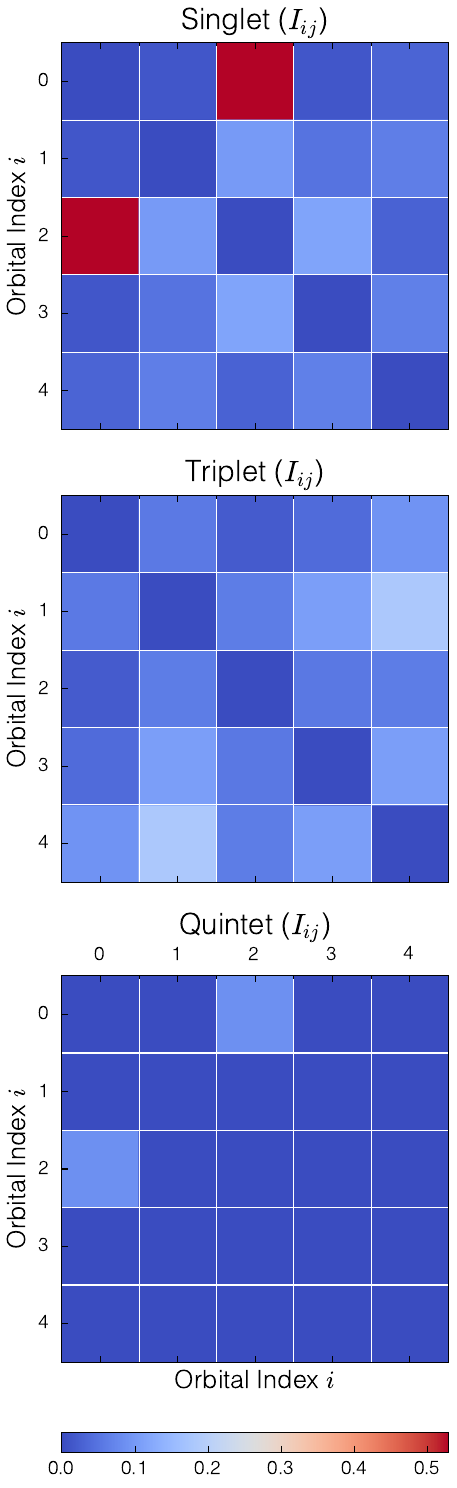}
		\label{fig:0.97_t3_mi}
	\end{subfigure}
	\begin{subfigure}[t]{0.16\textwidth}
		\caption{}
		\includegraphics[width=\textwidth]{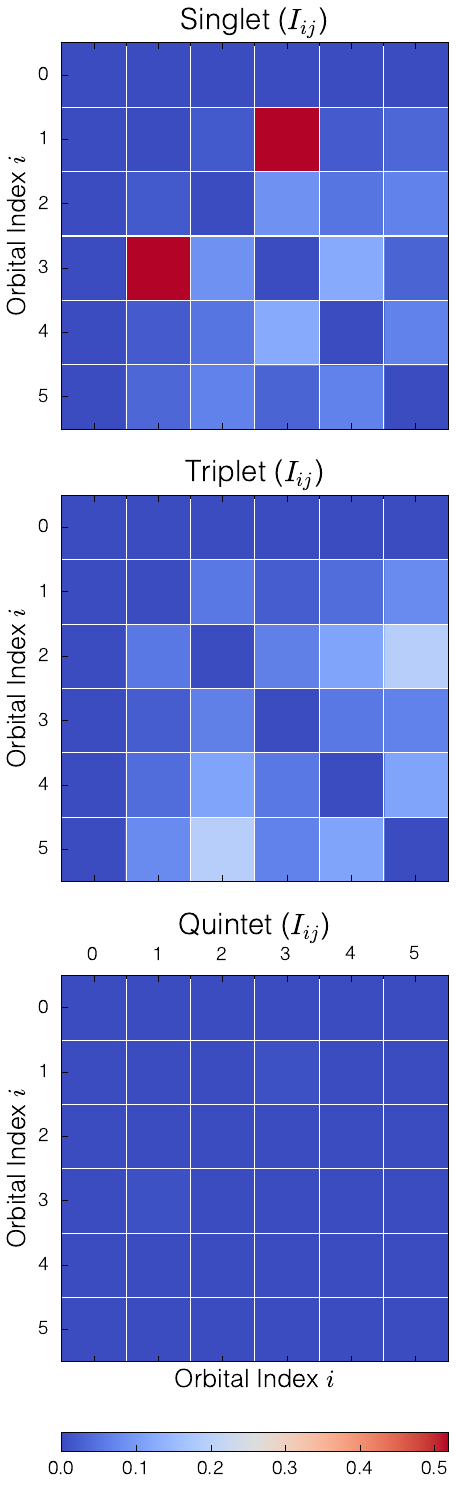}
		\label{fig:0.95_t3_mi}
	\end{subfigure}
	\begin{subfigure}[t]{0.16\textwidth}
		\caption{}
		\includegraphics[width=\textwidth]{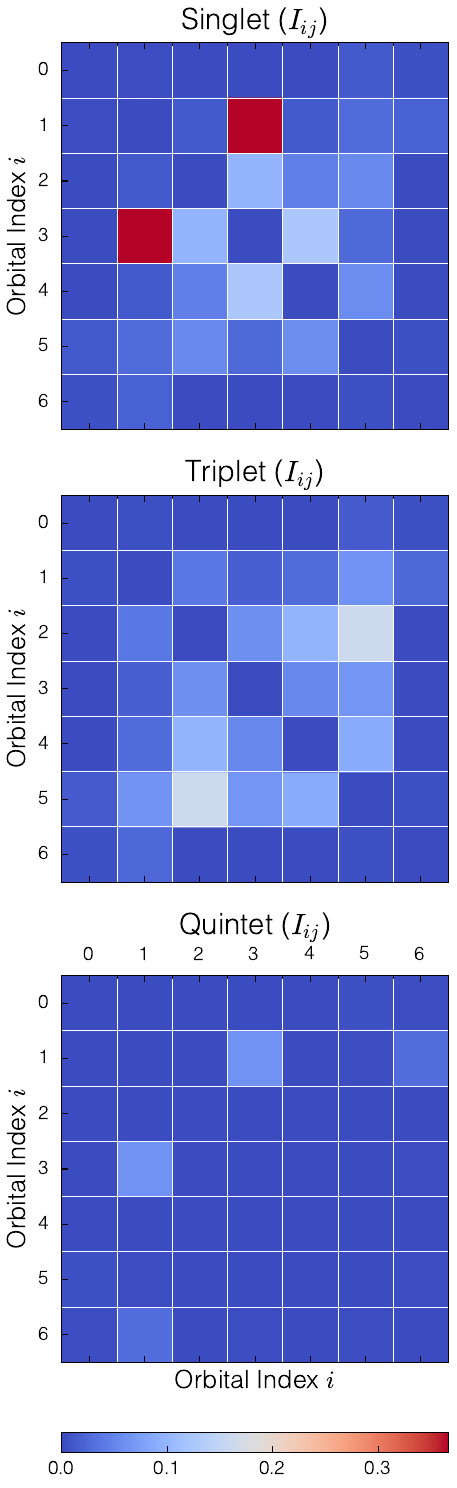}
		\label{fig:0.90_t3_mi}
	\end{subfigure}
	\begin{subfigure}[t]{0.16\textwidth}
		\caption{}
		\includegraphics[width=\textwidth]{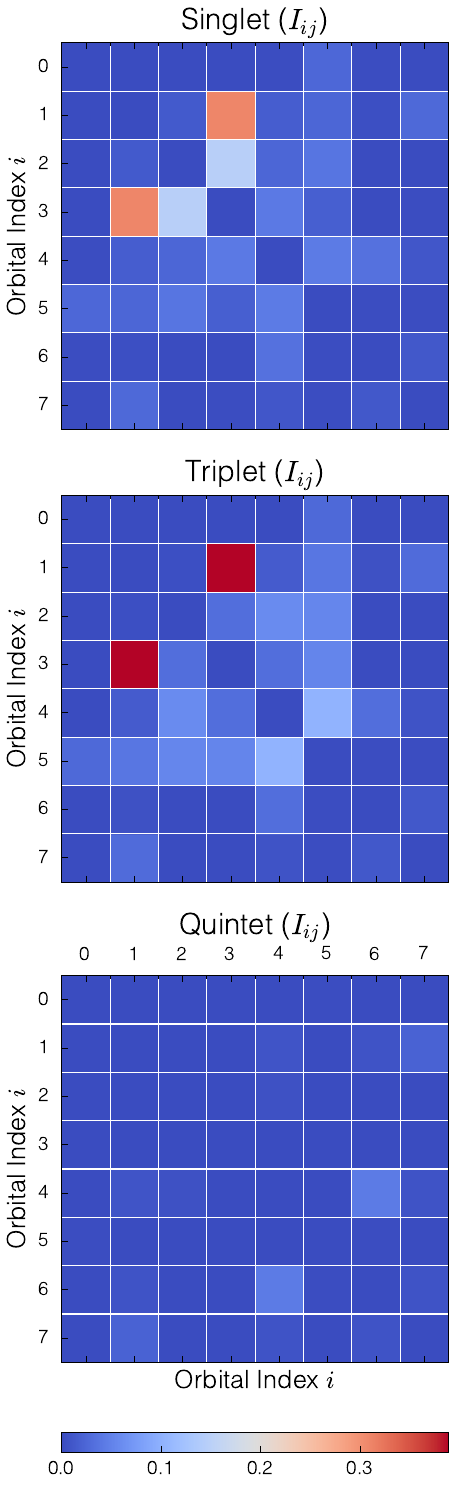}
		\label{fig:0.85_t3_mi}
	\end{subfigure}
	\begin{subfigure}[t]{0.16\textwidth}
		\caption{}
		\includegraphics[width=\textwidth]{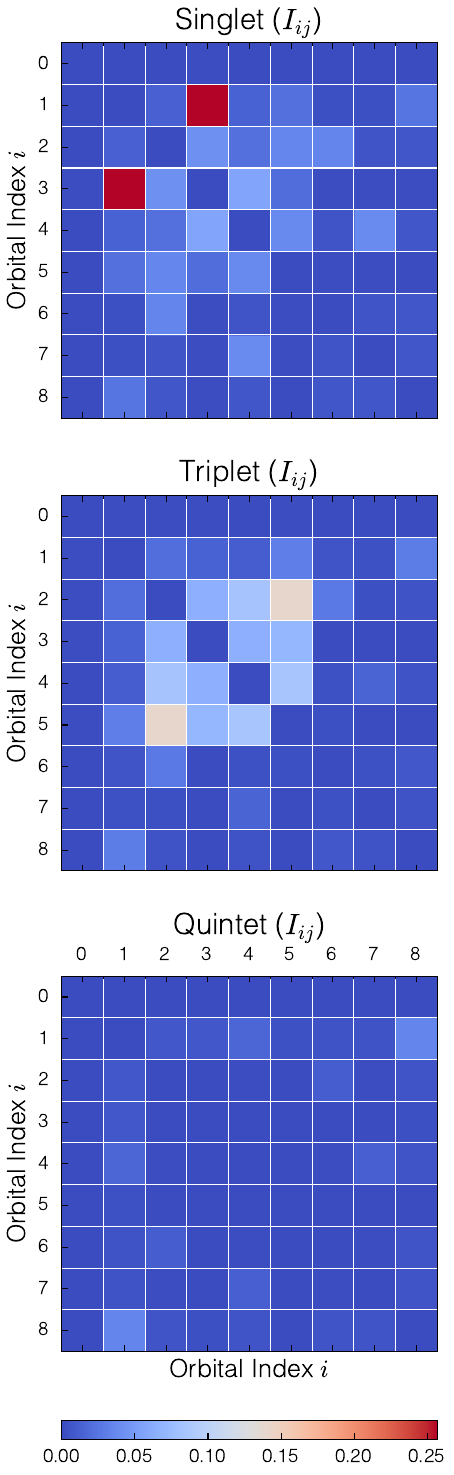}
		\label{fig:0.80_t3_mi}
	\end{subfigure}
	\begin{subfigure}[t]{0.16\textwidth}
		\caption{}
		\includegraphics[width=\textwidth]{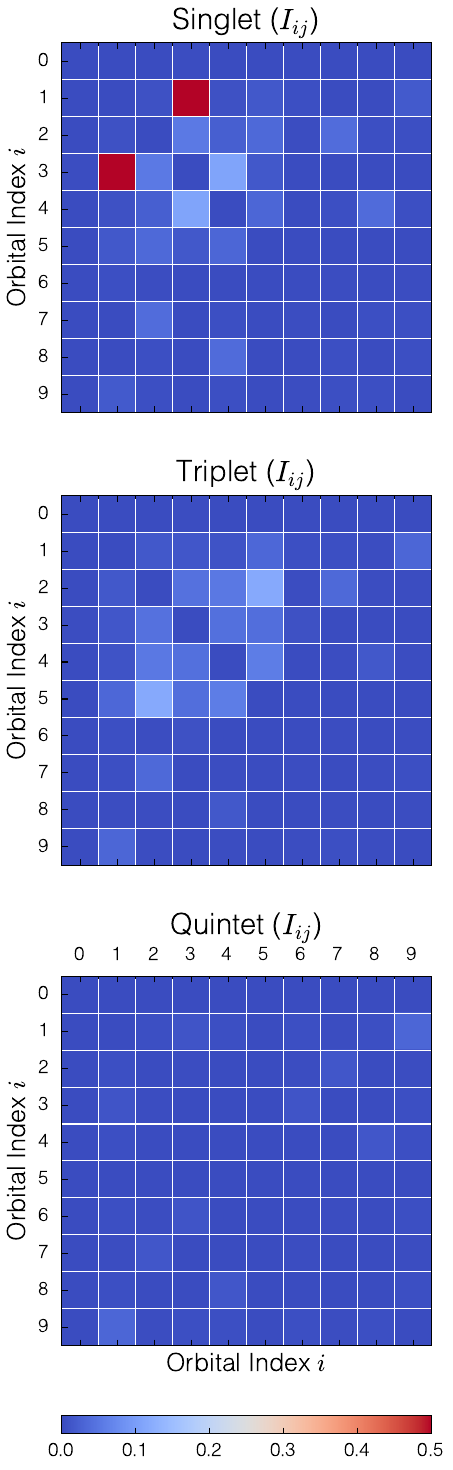}
		\label{fig:0.70_t3_mi}
	\end{subfigure}
	\caption{
		Orbital interactions for the T1 singlet $(S = 0)$, triplet $(S
		= 1)$ and quintet $(S = 2)$ spin states, respectively. The
		matrix plots show $I_{ij}$ for the final and converged spin
		states for the active spaces \textbf{(a)} (6e,5o), \textbf{(b)}
		(8e,6o), \textbf{(c)} (8e,7o), \textbf{(d)} (8e,8o),
		\textbf{(e)} (8e,9o) and \textbf{(f)} (8e,10o).
		\label{fig:t3_orbital_interaction}
	}
\end{figure*}

The orbital-pair mutual information $I_{ij}$ can be computed from the one- and
two-orbital entropies~\cite{Rissler2006,Boguslawski2013,Boguslawski2014}:
\begin{align}
	I_{ij} =  \frac{1}{2}(s_{i}(1) + s_{j}(1) - s_{ij}(2))(1 - \delta_{ij}) \geq 0,
\end{align}
\noindent
where $\delta_{ij}$ is the Kronecker delta, $s_{i}(1)$ is the one-orbital
entropy for orbital $i$, as in Eq.~\ref{eq:one_orbital_entropy}, and $s_{ij}(2)$
is the two-orbital entropy for an orbital pair $(i,j)$. The latter can be
calculated from the eigenvalues of the two-orbital reduced density
$\rho_{i,j}^{(2)}$ matrix for an orbital pair $(i,j)$:
\begin{align}
	s_{ij}(2) = -\sum_{\alpha=1}^{16} \omega_{\alpha,i,j} \ln\omega_{\alpha,i,j}.
\end{align}
\noindent
The non-zero matrix elements of the two-orbital reduced density matrix
$\rho_{i,j}^{(2)}$ can expressed in terms of the matrix elements of one-,two-,
three-, and four-particle reduced density matrices, see Tab.I in
Ref.~\cite{Boguslawski2014} for explicit formulae. We compute the matrix
elements for the one-, two-, three- and four-particle reduced density matrices
with final and converged state vectors for each spin state.
Fig.~\ref{fig:t0_orbital_interaction} and Fig.~\ref{fig:t3_orbital_interaction}
show the orbital-pair mutual information $I_{ij}$ for the T0 and T1 initial
states, respectively.
As mentioned in the main text, both singlet and triplet spin states show
significant orbital interactions while the quintet spin states have very small,
if any, orbital interactions. This behavior may be one of the reasons why the
state-averaged DMRGSCF total energies in Tab.~\ref{tab:dmrg_total_energies}
deviate from the state-averaged CASSCF total energies, even for such a large
bond dimension ($M = 10000$) used in the DMRG simulations.

\end{document}